# 1 Scholarly Twitter metrics

Stefanie Haustein
*stefanie.haustein@uottawa.ca*
[1] School for Information Studies, University of Ottawa, Ottawa, (Canada)
[2] Centre interuniversitaire de recherche sur la science et la technologie (CIRST), Université du Québec à Montréal, Montréal (Canada)

## Abstract
Twitter has arguably been the most popular among the data sources that form the basis of so-called altmetrics. Tweets to scholarly documents have been heralded as both early indicators of citations as well as measures of societal impact. This chapter provides an overview of Twitter activity as the basis for scholarly metrics from a critical point of view and equally describes the potential and limitations of scholarly Twitter metrics. By reviewing the literature on Twitter in scholarly communication and analyzing 24 million tweets linking to scholarly documents, it aims to provide a basic understanding of what tweets can and cannot measure in the context of research evaluation. Going beyond the limited explanatory power of low correlations between tweets and citations, this chapter considers what types of scholarly documents are popular on Twitter, and how, when and by whom they are diffused in order to understand what tweets to scholarly documents measure. Although this chapter is not able to solve the problems associated with the creation of meaningful metrics from social media, it highlights particular issues and aims to provide the basis for advanced scholarly Twitter metrics.

## 1.1 Introduction
Enabled by the digital revolution, the open access and open science movement, big data and the success of social media have shaken up the scholarly metrics landscape. Academic careers are no longer shaped only by peer-reviewed papers, citation impact and impact factors, university managers and funders now also want to know how researchers perform on social media and how much their work has impacted society at large.

Bibliometricians have started to adapt to the policy pull and technology push and expanded their repertoire of scholarly metrics to capture output and impact beyond the ivory tower, so far that some speak about a scientometric revolution (Bornmann, 2016). Metrics are no longer restricted to formal parts of communication but expand beyond the borders of the scholarly community (Cronin, 2013a). Similarly to how the Science Citation Index formed the field of bibliometric research and research evaluation, the altmetrics, or more precisely the social-media metrics landscape is being heavily shaped—if not entirely driven—by the availability of data, in particularly via Automated Programming Interfaces (APIs) (Haustein, 2016; Priem, 2014).

Twitter has arguably been at the epicenter of the earthquake that has shaken up the scholarly metrics landscape. The majority of altmetrics research has either focused on or included Twitter (see Sugimoto, Work, Larivière, & Haustein (2017) for a review of the literature). Following a general definition of scholarly metrics which include activity on social media (Haustein, 2016), scholarly Twitter metrics are defined as "indicators based on recorded events of acts [on Twitter] related to scholarly documents […] or scholarly agents […]" (Haustein, 2016, p. 416).

Although findings of an early study had suggested that tweets were a good early indicator of citations for papers published in the *Journal of Medical Internet Research* (Eysenbach, 2011), the generalizability of this claim was refuted by low correlations reported by more representative studies (Costas, Zahedi, & Wouters, 2015a; Haustein, Peters, Sugimoto, Thelwall, & Larivière, 2014). Low correlations between tweets and citations did, however, spark hopes that Twitter activity was able to reflect impact on users and use beyond citing authors—a new type of previously unmeasurable impact, possibly on society at large. Twitter's popularity in the altmetrics realm has essentially been caused by two factors, which are both heavily influenced by technology and the data push and policy pull described above:

a) a significant number of scholarly articles are shared on Twitter, producing a measurable signal;
b) Twitter is a social media platform created for and used by a wide and general user base, which theoretically has the potential to measure impact on society at large.

As tweets represent an "unprecedented opportunity to study human communication and social networks" (G. Miller, 2011, p. 1814), Twitter is being used to analyze a variety of social phenomena. Centering on either the message (i.e., the tweet, its content and associated metadata) or social connections (i.e., the network of follower-followee relations), tweets have been used to show discussions during upcoming elections, how people communicate during natural disasters, political upheaval, cultural events and conferences and have even been used to predict elections outcomes and the stock market (Rogers, 2014; Weller, Bruns, Burgess, Mahrt, & Puschmann, 2014).

While only a small share of academics use Twitter for scholarly communication (Rowlands, Nicholas, Russell, Canty, & Watkinson, 2011; Van Noorden, 2014) and to diffuse scientific publications (Priem & Costello, 2010), more than one fifth of recent journal articles are being tweeted (Haustein, Costas, & Larivière, 2015), which adumbrates that it is non-academics who engage with scholarly publications on Twitter. At this point, social media-based indicators have flourished rather as vanity measures—culminating in a tongue-in-cheek metric called the Kardashian Index (Hall, 2014)—than as validated indicators of societal impact. Even though altmetrics have left their mark on the scholarly publishing and metrics landscape (Desrochers et al., 2018), they have not (yet) established themselves within the reward system of science, where citations remain the only hard currency:

> Neither Twitter mentions nor Facebook 'likes' are, for now at any rate, accepted currencies in the academic marketplace; you are not going to get promoted for having been liked a lot, though it may well boost your ego. (Cronin, 2013b, p. 1523)

Still, almost all big publishing houses now report some form of article-level metric based on social media activity, including tweets. Despite the lack of validation and a clear definition regarding the type of impact measured, the number of tweets are thus already used as scholarly metrics "in the wild" (Priem, Piwowar, & Hemminger, 2012).

This chapter aims to contribute to the understanding of Twitter and Twitter-based metrics with a particular focus on their potential and limitations when applied as scholarly metrics. To provide some context for the meaning of scholarly impact measures derived from tweets and Twitter activity, this chapter describes Twitter's role in scholarly communication. It depicts how Twitter is used in academia and how scholarly contents are diffused and discussed in tweets. The chapter provides an overview of the literature of Twitter use by the scholarly community and scholarly output on Twitter. The latter part is supported by empirical results based on an analysis of 24 million tweets mentioning scientific papers captured by the data provider Altmetric.com. Both the review of the relevant literature and the patterns extracted from the Twitter data are intended to contribute to the understanding of what type of scholarly contents are diffused on Twitter, who is diffusing them, when and how. This will help to assess Twitter metrics as valid impact indicators and to interpret their meaning.

## 1.2 Twitter in scholarly communication

Twitter launched in 2006 as a public instant messaging service and evolved from an urban lifestyle social network, where users would update their friends about what they were doing, to a platform for communicating news and events used by 500 million users worldwide, or 23% of US adults online (Duggan, Ellison, Lampe, Lenhart, & Madden, 2015). Although other microblogging platforms (e.g., Sina Weibo, tumblr, Plurk) exist, tweeting has become a synonym (and preferred term) for microblogging and Twitter the most popular service. Twitter constraints microposts to a maximum length of 140 characters, a restriction that originates from the 160 character limit of text messages. Users can follow each other and create user lists to manage the updates they receive from other Twitter users. Similarly to regular blogs, microblogs are ordered sequentially in reverse chronological order and, due to their brevity, usually appear more frequently (Puschmann, 2014), making Twitter the "most dynamic and concise form of information exchange on social media" (Grajales, Sheps, Ho, Novak-Lauscher, & Eysenbach, 2014, p. 5). While the brevity of tweets is seen as a restriction by some, others perceive it as a particular advantage:

> The brevity of messages allows [tweets] to be produced, consumed, and shared without a significant amount of effort, allowing a fastpaced conversational environment to emerge. (boyd, Golder, & Lotan, 2010, p. 10)

Tweets have three major specific affordances, which facilitate communication on the platform: retweets (RTs), user mentions (@mentions) and hashtags (keywords following #). All of these functions originated within the Twitter user base and were eventually adopted by Twitter, representing a co-creation of functions by users and developers. Twitter provides three main levels of communication: interpersonal communication on the micro level, meso-level exchanges of people who are directly connected through their network of followers and followees, and hashtag-centered macro-level communication which enables exchanges among all Twitter users with common interests (Bruns & Moe, 2014).

In academia Twitter is used to disseminate and discuss scholarly outputs and other relevant information; maintain collaborations or find new ones; as a virtual "water cooler" (Veletsianos, 2012, p. 347) for social networking with colleagues; to increase student participation in teaching; as a backchannel at scientific conferences to foster discussions among conference attendees and those who participate remotely; as well as to increase visibility and reach wider audiences (Nentwich, 2011; Osterrieder, 2013; Pearce, Weller, Scanlon, & Ashleigh, 2010; Van Noorden, 2014; Veletsianos, 2012; Zhao & Rosson, 2009).

### 1.2.1 Twitter uptake

Twitter is used by various stakeholders in the scholarly community, including individual researchers and academics, journals and publishers, universities and other academic institutions, as well as at scholarly conferences. From the perspective of using Twitter activity as the basis for scholarly metrics, it is essential to know Twitter uptake in academia, as it informs about biases and differences between disciplines and other user demographics, which may have a direct effect on derived metrics.

#### *1.2.1.1 Scholars on Twitter*

In the scholarly context, Twitter use by academics lags behind its uptake among the general public. Although the majority of researchers are aware of the platform, most do not make use of it in a professional context, giving it the reputation of a hype medium in academia (Carpenter, Wetheridge, Tanner, & Smith, 2012; Gu & Widén-Wulff, 2011; Pscheida, Albrecht, Herbst, Minet, & Köhler, 2013; Van Noorden, 2014). A certain reluctance in academia to use Twitter might be caused by its perception as a shallow medium that is used to communicate "pointless babble" (Kelly, 2009, p. 5) rather than informative content (Rogers, 2014). Described as "phatic" (V. Miller, 2008, p. 396), Twitter is less about what people tweet rather than how they are connected.

Reported Twitter uptake shows extreme variations depending on user demographics—in particular disciplinary orientation—and when a study was conducted; it usually stays behind use of other social media. For example, a survey among 2,414 researchers conducted in 2010 demonstrated that while more than three quarters used social media, less than one fifth were on Twitter (Rowlands et al., 2011). A more recent study conducted by *Nature* also showed that Twitter was among the least used social media platforms in academia: while almost half of 3,027 science and engineering researchers regularly used ResearchGate, only 13% regularly visited Twitter (Van Noorden, 2014). At almost one quarter of regular users, Twitter uptake was higher among the 482 social science and humanities scholars participating in the same survey (Van Noorden, 2014).

Depending on the sample and when the survey was conducted, Twitter uptake varied heavily between a few percent to more than one third of surveyed scholars using Twitter, which calls the representativeness of findings into question. Moreover, surveys vary in terms of whether or not they differentiate between general Twitter uptake, Twitter use for scholarly communication and professional purposes or active vs. passive use, which further complicates comparison and generalization of findings regarding Twitter uptake in academia. Twitter represented the social media tool with the highest difference between awareness and use. Although known by 97% of university staff in Germany, as few as 15% used Twitter and 10% used in a professional context (Pscheida et al., 2013). A similar use-to-awareness ratio was found by other studies, for example, at Finnish universities (Gu & Widén-Wulff, 2011) or researchers surveyed by *Nature* (Van Noorden, 2014), and academic staff in Germany (Weller, Dornstädter, Freimanis, Klein, & Perez, 2010).

Most studies found self-reported Twitter use in academia at around 15%; an uptake of 13-16% was reported for surveys based on 215 health services and policy researchers (Grande et al., 2014), 454 geographers (Wilson & Starkweather, 2014), 1,058 UK academic staff (Tenopir, Volentine, & King, 2013) and 3,027 scientists and engineers (Van Noorden, 2014), while uptake was lower (7-10%) for academics in Germany (Weller et al., 2010) and the UK (Procter et al., 2010; Rowlands et al., 2011). Although 18% used it, Twitter was the least popular social media tool of 345 European scholars (Ponte & Simon, 2011). The highest Twitter use was reported for a survey of 126 Finnish university staff at 23% (Gu & Widén-Wulff, 2011), 1,910 professors at US universities at 32%—23% for professional purposes—(Bowman, 2015b), 382 urologists attending a conference at 36% (Loeb et al., 2014) and 71 participants of the 2010 *Science & Technology Indicators* (STI) conference in Leiden at 44% (Haustein, Peters, Bar-Ilan, et al., 2014).

As an alternative to determining Twitter uptake through self-reported use in surveys, studies have also assessed the extent of scholarly microblogging based on Twitter activity of scholars. Identifying scholars on Twitter is challenging, as the 160-character Twitter bio and the provided user name are often the only basis for identification. Most studies thus search for Twitter users based on a list of names of academics (Darling, Shiffman, Côté, & Drew, 2013; Desai et al., 2012; Desai, Patwardhan, & Coore, 2014; Priem, Costello, & Dzuba, 2012; Work, Haustein, Bowman, & Larivière, 2015) or apply snowball sampling starting from a set of known scholars on Twitter (Chretien, Azar, & Kind, 2011; Holmberg, Bowman, Haustein, & Peters, 2014; Holmberg & Thelwall, 2014). Searching for a list of 8,038 US and UK university staff, Priem, Costello, and Dzuba (2012) found Twitter accounts for 2.5% of them. Although the authors admit that their study underestimated Twitter use, it reflected that the microblogging platform is not popular in academia and thus confirms finding by most surveys. Investigating Twitter use in the scientometric community, Bar-Ilan et al. (2012) found Twitter profiles for 9 of the 57 presenters at the 2010 STI conference.

Other than searching for known scholars on Twitter, some studies try to extract information from Twitter to identify scholarly users. The most common approach is to classify users based on searching for specific words in the Twitter bio. Retrieving users whose Twitter bio contained words such as *university*, *PhD* or *professor*, Barthel et al. (2015) identified scientists' Twitter accounts with a precision of 88%. False positives contained university accounts or those of non-academic staff at research institutions. Recall cannot be determined in such studies as the number of false negatives, that is the scientists on Twitter who do not include any of the queried keywords in their self-descriptions, remains unknown.

Altmetric also applies a keyword-based approach to categorize Twitter users as *scientists*, *science communicators (journalists, bloggers, editors)*, *practitioners (doctors, other healthcare professionals)* and *members of the general public*. It should be noted that Altmetric's *general public* category includes all users that can not be classified as belonging to any of the other three groups and is therefore not a good indicator of how much an article has been tweeted by members of the general public. An obvious limitation of the keyword approach is that it is unable to capture scholars who do not identify themselves as such or who do not use the terminology or language covered by the list of keywords. However, many scholars seem to reveal their professional personas on Twitter. Ninety percent of doctoral students funded by the Canadian Social Sciences and Humanities Research Council (SSHRC) identified as academics on Twitter (Work et al., 2015), 87% of surveyed US university professors claimed to mention both their professional title and place of work in their Twitter profiles (Bowman, 2015b) and 78% of Twitter users who self-identified as a physicians used their full names (Chretien et al., 2011). This willingness to reveal their scholarly identities on Twitter suggests that scholars make use of the microblogging platform in a professional context at least to some extent.

Ke, Ahn, and Sugimoto (2017) took advantage of crowdsourced Twitter lists to identify scholars. Based on a method introduced by Sharma, Ghosh, Benevenuto, Ganguly, and Gummadi (2012), they identified scientists on Twitter with an approach based on membership in scientific Twitter user lists. Other studies have estimated Twitter activity by scholars by analyzing users who engage with scholarly content on Twitter. Hadgu and Jäschke (2014) applied machine learning to automatically identify scholars on Twitter based on a training set of users whose tweets contained a computer science conference hashtag, while others selected users who have tweeted scientific papers (Alperin, 2015b; Haustein, Bowman, & Costas, 2015a; Haustein & Costas, 2015b; Holmberg & Thelwall, 2014; Tsou, Bowman, Ghazinejad, & Sugimoto, 2015). Since the latter type of studies focuses on categorizing who is tweeting about scholarly contents rather than estimating Twitter uptake in academia, these studies are discussed in more detail below.

#### *1.2.1.2 Scientific conferences*

Twitter does particularly well in fostering communication among people participating in shared experiences (Rogers, 2014), which may be why tweeting at scholarly conferences has been one of the earliest and most popular uses of Twitter in academia. Almost every scientific conference today has a specific hashtag to connect attendees and those interested but not able to attend in person, thus expanding the conference audience to include remote participants (Bonetta, 2009; Sopan, Rey, Butler, & Shneiderman, 2012; Weller, Dröge, & Puschmann, 2011). Apart from increasing the visibility of presentations, tweeting at scientific conferences has introduced another level of communication, creating backchannel discussions online among participants complementing presentations and discussions which take place at the meeting. Conference tweets usually directly refer to presentations and discussions during sessions and sometimes summarize key take-away points (Chaudhry, Glode, Gillman, & Miller, 2012; McKendrick, Cumming, & Lee, 2012; Mishori, Levy, & Donvan, 2014). Other motivations to tweet at a scientific conferences were to share information and learn about discussions in parallel sessions, networking with others and feeling a sense of connectedness, as well as note-taking (McKendrick et al., 2012). A significant number of tweets associated with two medical conference were uninformative or promotional (Cochran, Kao, Gusani, Suliburk, & Nwomeh, 2014; Desai et al., 2012).

Due to the ease of collecting tweets with a particular hashtag, as well as Twitter's relative popularity in the context of scientific conferences, there are countless studies analyzing scholarly Twitter use based on tweets with conference hashtags (Chaudhry et al., 2012; Cochran et al., 2014; Ferguson et al., 2014; Hawkins, Duszak, & Rawson, 2014; Hawkins et al., 2014; Jalali & Wood, 2013; Letierce, Passant, Breslin, & Decker, 2010; McKendrick et al., 2012; Mishori et al., 2014; Reinhardt, Ebner, Beham, & Costa, 2009; Sopan et al., 2012; Weller et al., 2011).

Similar to the overall uptake among scholars, Twitter activity at conferences differs among disciplines as well as individual conferences and increased over the years. Overall, only a small share of conference participants contributed to discussions on Twitter: Less than 2% of attendees of the *American Society of Nephrology*'s 2011 conference (Desai et al., 2012) and less than 3% of participants of the 2012 *Winter Scientific Meeting of the Association of Anesthetists of Great Britain and Ireland* tweeted (McKendrick et al., 2012). Another medical conference in 2013 reported higher Twitter engagement, as 13% of conference attendees tweeted using the conference hashtag (Mishori et al., 2014). Longitudinal studies also observed an increase in Twitter activity at conferences over the years (Chaudhry et al., 2012; Hawkins et al., 2014; Mishori et al., 2014). For example, 2% of conference participants tweeted at the *2010 Annual Meeting of the American Society of Clinical Oncology*, while 5% contributed to conference tweets in 2011. Similarly, the number of tweets nearly doubled from 4,456 tweets in the first to 8,188 in the following year (Chaudhry et al., 2012). A similar increase was observed for the 2011 and 2012 annual meetings of the *Radiological Society of North America* (Hawkins et al., 2014). Conference-related discussions on Twitter are not restricted to in-person attendees. In fact, at some conferences the majority of Twitter users only participate remotely (Sopan et al., 2012).

Just as with other social media and information in general, tweeting activity is usually heavily skewed with a few users contributing the majority of tweets at conferences (Chaudhry et al., 2012; Cochran et al., 2014; Mishori et al., 2014). Tweeting about a conference has been shown to lead to an increase in the number of followers regardless of attending in-person or remotely. Follower counts grew particularly for speakers and in-person attendees, while the number of followers grew least for remote participants (Sopan et al., 2012). Most organizers of scientific conferences embrace the potential of increasing visibility and outreach and thus encourage tweeting through a conference-specific hashtag. Some also specifically display conference-related tweets in real time and thus make tweeting activity visual to participants who are not on Twitter (Ferguson et al., 2014; Jalali & Wood, 2013; Sopan et al., 2012; Weller et al., 2011).

#### *1.2.1.3 Journals and publishers*

Twitter's technological features afford direct connections and two-way conversations between users changing what was traditionally known as a unidirectional sender-audience relationship. Opposed to traditional publishing and mass media, Twitter has given rise to personal publics of audiences (Schmidt, 2014). This direct link between the sender and receiver has changed the relationship with audiences; for example, musicians use Twitter to market their own brand and respond to @replies from fans to seek out in-person interactions (Baym, 2014). TV audiences turn Twitter into a virtual lounge room when they connect with other users discussing TV events in real time (Harrington, 2014). Similarly, discussions of scientific publications can now happen publically, when readers share their opinions on Twitter. A specific use case are Twitter journal clubs, an adaption of small-group in-person journal clubs that are common particular in the medical sciences (Leung, Siassakos, & Khan, 2015; Mehta & Flickinger, 2014; Thangasamy et al., 2014; Topf & Hiremath, 2015; Whitburn, Walshe, & Sleeman, 2015). Twitter journal clubs are used to discuss and review recent publications and educate researchers and practitioners; in the medical sciences they also have the advantage over their offline predecessors to directly involve patients (Mehta & Flickinger, 2014). Often these journal clubs are initiated or at least supported by journals to promote their publications. A journal club initiated by a gynecology journal showed that discussing papers and making them freely available has boosted their Altmetric scores (Leung et al., 2015). Twitter journal clubs also motivated authors of discussed papers to create Twitter accounts (Thangasamy et al., 2014).

With journals and authors on Twitter, readers can get in touch directly and involve them in discussions using @mentions, tearing cracks in the wall of traditional gatekeeping, as "Twitter makes it possible to directly connect journal readers at various stages of training with authors and editors" (Mehta & Flickinger, 2014, p. 1317). Many journals and publishers have started to use Twitter as a marketing instrument to increase online visibility and promote published contents. These accounts can be used to create a personalized audience relationship and to foster interaction among readers. Similar to the mix of professional and personal interactions by academics on Twitter, the lines between scholarly communication and marketing campaign are blurred for accounts maintained by journals and publishers. Almost half of the 25 general medicine journals with the highest impact factor in 2010 had a Twitter presence (Kamel Boulos & Anderson, 2012), while Twitter uptake was lower for other sets of journals: 24% of 33 urology journals (Nason et al., 2015), 2 of the top 10 ophthalmology journals (Micieli & Micieli, 2012), 16% of 100 *Web of Science* (WoS) journals (Kortelainen & Katvala, 2012) and 14% of 102 journals specialized in dermatology (Amir et al., 2014) maintained an account. As most of these studies focused on the top journals according to the journal impact factor, Twitter uptake might be biased towards high-impact journals and slightly lower when including others. The variation suggests similar differences between disciplines as observed for Twitter uptake by individual scholars and conferences.

While most journal accounts are used to share articles and news (Zedda & Barbaro, 2015) and often tweet the article title (Friedrich, Bowman, Stock, & Haustein, 2015; Thelwall, Tsou, Weingart, Holmberg, & Haustein, 2013), some journals have incorporated tweeting into the formal communication process. In addition to regular abstracts, they ask authors to write so-called tweetable abstracts that meet the 140-character restrictions, which are used to attract readers on Twitter (Darling et al., 2013). Twitter even interfered with the journal's role in scholarly communication, when a genomics paper was criticized and corrected results posted in a tweet, leading to a conflict with the authors of the criticized paper (Woolston, 2015).

Even if a journal is not represented by a proper Twitter account, it is likely that its publisher is. Zedda and Barbaro (2015) found that Twitter adoption was particularly common among 76 publishers in science, technology and medicine; 89% had official Twitter accounts, exceeding the presence on any other social media platform, and 74% had embedded tweet buttons that allowed readers to directly share publications on Twitter. Promotion of publications by publishers seem to be welcomed by authors, as a survey by *Nature Publishing Group* revealed that almost one fifth of authors would consider it a very valuable service if publishers promoted papers using marketing and social media (Nature Publishing Group, 2015).

As shown in the analysis of Twitter accounts diffusing scientific articles below, accounts maintained by journals and publishing houses are responsible for a significant amount of tweets mentioning scientific articles. Once Twitter metrics are being used to evaluate journal impact, these types of self-tweets might be considered as a type of *gaming* in a manner similar to journal self-citations and citation cartels to boost the impact factor (Seglen, 1997; Van Noorden, 2013). With publishers invested in the success of their journals, *tweet cartels* and *tweet stacking* in analogy to their citation equivalents are easily conceivable and even easier to implement. While the WoS excludes journals from the *Journal Citation Reports*, which have been caught increasing their impact factors artificially, companies like Altmetric.com and Plum Analytics do not (yet) intervene in such self-promotional activity.

#### *1.2.1.4 Universities and academic libraries*

Scholarly institutions are affected by Twitter's impact on academia on two levels: they exploit the microblogging platform to increase their visibility (and that of their members) and provide guidelines and recommendations for their members to navigate the new communication space. The Association of American University Professors updated their report on *Academic Freedom and Electronic Communication* (Association of American University Professors, 2013) in reaction to a university rescinding a tenure-track job offer to an English scholar who had made an anti-Semitic comment on Twitter (Herman, 2014). The updated report emphasized that professors enjoy academic freedom even when they comment on social media and particularly addressed the blurring of boundaries between private and professional opinions on social media. It stressed how, in this new context, comments are particularly prone to be misunderstood and misinterpreted, as they are often taken out of context:

> Electronic communications can be altered, or presented selectively, such that they are decontextualized and take on implicit meanings different from their author's original intent. With the advent of social media such concerns about the widespread circulation and compromised integrity of communications that in print might have been essentially private have only multiplied further. (Association of American University Professors, 2013, p. 42)

The report further recommends that universities and other academic institutions, along with their staff, develop policies that address the use of social media. In general, academic institutions lack specific social media guidelines or address social media in policies. Although more and more institutions adopt specific policies (Pasquini & Evangelopoulos, 2015), only half of US doctorate-granting universities had a social media policy, while rates were even lower for other universities and colleges. At the same time, Twitter was specifically mentioned in more than 80% of policies (Pomerantz, Hank, & Sugimoto, 2015).

The majority of university Twitter accounts apply a so-called megaphone model of communication, where news and information concerning the institution are broadcasted following a traditional communication model (Kimmons, Veletsianos, & Woodward, 2017; Linvill, McGee, & Hicks, 2012). Universities use Twitter for public relations, dissemination of news and events, as well as recruitment (Kimmons et al., 2017). Ninety-four percent of 474 US university admission officers reported that their institution had a Twitter account (Barnes & Lescault, 2013) and 96% of the websites of 100 US colleges linked to Twitter (Greenwood, 2012). On the departmental level, it was less common to be represented with an organizational account, as only 8% of 183 US radiology departments had a Twitter presence (Prabhu & Rosenkrantz, 2015). Twitter was also commonly used for faculty development at medical schools (Cahn, Benjamin, & Shanahan, 2013). Analyzing the Twitter activity of 29 Israeli universities and colleges, Forkosh-Baruch and Hershkovitz (2012) found significant differences between both types of institutions. Colleges were more likely than universities to post social tweets. While almost half of the tweets by universities focused on research conducted elsewhere, colleges focused more on reporting the work by its own researchers.

Twitter was also frequently used by academic libraries as a marketing instrument, to communicate with patrons, to announce new resources and promote services (Boateng & Quan Liu, 2014; Hussain, 2015; Shulman, Yep, & Tomé, 2015). On par with Facebook at a 63% adoption rate, Twitter was the most commonly used social media platform among 38 surveyed academic libraries from different countries (Chu & Du, 2013), while all of the 100 US university libraries analyzed by Boateng and Quan Liu (2014) maintained a Twitter presences. In Canadian academic libraries Twitter adoption was lower at 47% (Verishagen & Hank, 2014).

### 1.2.2 *Twitter use*

Apart from Twitter uptake, scholarly Twitter metrics are further influenced by *how* Twitter is used. Academics use social media to share information, for impression management and to increase their visibility online, to network and establish a presence across platforms, to request and offer help, expand learning opportunities, or simply to be social (Veletsianos, 2012). Twitter specifically was used mostly to tweet work-related content, discover peers working on similar research, follow research-related discussions and get recommendations for papers (Van Noorden, 2014).

One central motivation for scholars to tweet is to communicate and explain their work to lay people. As many science communicators are active on Twitter, they help to bridge the gap between the scholarly community and the general public. Science communicators were the largest user group of 518 Twitter users mentioned in tweets by 32 astrophysicists (Holmberg et al., 2014). An evolutionary biology professor valued Twitter to communicate his work to the general public:

> Twitter and regular blogging are more effective than anything else I do to publicize a paper, which was really surprising to me […]. If you do it right, Twitter is an effective way of telling people about your work. (Bonetta, 2009, p. 453)

However, most researchers still preferred traditional media over Twitter to promote their research (Wilkinson & Weitkamp, 2013).

Even when identifying professionally on Twitter, a large share of tweets by scholars are not related to their work or academia in general (Bowman, 2015b; Haustein, Bowman, Holmberg, Peters, & Larivière, 2014; Mou, 2014; Pscheida et al., 2013; Van Noorden, 2014). The *Nature* survey found that 21% of scientists who used Twitter regularly did not use it professionally and 28% said that they never posted content about their work (Van Noorden, 2014). Bowman (2015a, 2015b) reported that, while 29% of American university professors used Twitter strictly in a professional way and 42% used it for both for personal and professional reasons, the vast majority of tweets were coded as personal (78%) rather than professional (19%). Again, large variations can be observed between disciplines as well as individual Twitter users (Chretien et al., 2011; Holmberg & Thelwall, 2014; Loeb et al., 2014; Mou, 2014; Priem, Costello, et al., 2012; Work et al., 2015). Examining more than half a million tweets from 447 researchers, Holmberg and Thelwall (2014) found that less than 4% of tweets were classified as scholarly communication and results varied between disciplines ranging from less than 1% for sociology up to 34% for biochemistry. A study on emergency physicians' tweeting behavior found that 49% of their tweets were related to health or medical issues, 21% were personal, 12% self-promotional and 3% considered unprofessional as they contained profanity, were discriminatory or violated patient privacy (Chretien et al., 2011). In a sample of tweets by funded doctoral students in the social sciences and the humanities in Canada, 4% of tweets were related to their thesis, 21% to the discipline and 5% to academia in general, while 70% of tweets were coded as non-academic (Work et al., 2015). Personal use also prevailed among 382 urologists (Loeb et al., 2014).

These findings highlight that even when scholars identify professionally on Twitter and use the platform for scholarly purposes, many tweets will be irrelevant to scholarly communication and should thus be excluded from a scholarly indicator perspective:

> The lack of a dividing line between scientists and non-scientists, as well as the great variety of topics that even scientists tweet about mean that Twitter is not comparable to the orderly world of science publishing, where every piece of information is assumed to be relevant. Instead, a typical user's timeline is likely to be populated both by scholarly content and personal remarks, more or less side by side. (Puschmann, 2014, p. 98)

Due to their brevity and the fact that when analyzing tweets they are often taken out of context, categorizing tweet content is as difficult as it is to classify Twitter users (Bowman, 2015b). Distinguishing between scientific and non-scientific tweets is especially challenging (Holmberg & Thelwall, 2014).

Large variations can also be found between individual tweeters in terms of how often they tweet. A group of astrophysicists analyzed by Haustein et al. (2014) tweeted, on average, between 0 and 58 times per day. Tweets to scientific papers have been shown to peak shortly after their publication and decay rapidly within just a few days. For example, 80% of *arXiv* submissions received the largest number of tweets the day after they were published (Shuai, Pepe, & Bollen, 2012). Similarly tweeted half-life was 0 days for papers published in the *Journal of Medical Internet Research* (Eysenbach, 2011) and 39% of a sample of tweets linking to a scholarly document referred to those published within one week before (Priem & Costello, 2010). Determining the delay between publication and first tweet as well as half-lives on Twitter is, however, challenging due to the ambiguity of publication dates (Haustein, Bowman, & Costas, 2015b).

Since not all Twitter use culminates in a tweet, a large share of activity remains invisible and thus unmeasurable. In fact, passive use prevailed among UK doctoral students on Twitter (Carpenter et al., 2012) and seems to be common for scholarly use of social media in general. While most academics access and view information, only a minority actively contributes by creating content on social media (Procter et al., 2010; Tenopir et al., 2013); less than 2% of 1,078 UK researchers surveyed actively contributed daily (Tenopir et al., 2013).

#### *1.2.2.1 Tweeting links*

The most frequent use of Twitter among researchers in higher education was to share information, resources or media (Veletsianos, 2012). A survey among US university professors revealed that embedding URLs was the most commonly used Twitter affordance. Half of the survey participants claimed to tweet links either sometimes, mostly or always (Bowman, 2015b). Links are a common way to send more information than 140 characters would fit. Addressing the length limitation, a scholar explained:

> It is a double-edged sword. The majority of my tweets are pointers to other resources, so there is a headline—an enticement in other words—and a link to the resource. You don't need more than 140 characters for that. (Bonetta, 2009, p. 453)

Weller and Puschmann (2011) refer to links in tweets as "external citations" (Weller & Puschmann, 2011, p. 2). Studies about scholars on Twitter show that they make frequent use of tweeting URLs, as the share of tweets with links exceeds that observed for general Twitter users (boyd et al., 2010; Harrington, 2014). About one third of 68,232 tweets sent by 37 astrophysicists (Haustein, Bowman, Holmberg, et al., 2014) and 38% of 22,258 tweets posted by Canadian social sciences and humanities doctoral students contained links (Work et al., 2015). Tweeting links was even more popular among scholars studied by Weller and Puschmann (2011) and emergency physicians analyzed by Chretien et al. (2011), as respectively 55% and 58% of their tweets contained URLs. Links were much more common when a sample of 445 US professors tweeted professionally: 69% of professional tweets contained URLs, while only 15% of personal tweets did (Bowman, 2015b). Tweets with the #www2010 and #mla09 conference hashtags linked to a website in 40% and 27% of the cases, repeating each unique URL less than three times (Weller et al., 2011).

Priem and Costello (2010) found that 6% of a sample of 2,322 of tweets by academics containing a URL mentioned a scholarly publication, 52% of which were first-order and 48% second-order links (i.e., via another website) to the document. Similarly, Holmberg and Thelwall (2014) found that scholarly tweets frequently contained a link to scholarly publications via a blog post about the paper.

First-order links were significantly more likely to refer to open access articles (Priem & Costello, 2010). Tweets containing the #iswc2009 conference hashtag linked to applications (e.g., online services or research projects; 31%), the conference website (21%), blog posts (12%), slideshows (12%) and publications (9%) (Letierce et al., 2010). Blogs were the most common linked resources in conference tweets analyzed by Weller and colleagues (2011), while news websites were a frequent link destination of tweets sent by Canadian doctoral students, even when discussing scholarly topics (Work et al., 2015).

When linking to scholarly papers, tweets often contained the paper title and rarely expressed any recommendation or sentiment (Friedrich, Bowman, Stock, et al., 2015; Thelwall et al., 2013). The great majority of tweeted articles were published very recently (Eysenbach, 2011; Holmberg & Thelwall, 2014; Priem & Costello, 2010). According to surveys asking for motivations to use social media, finding relevant publications and staying up-to-date with the literature was found to be a frequent, yet passive, use of Twitter (Van Noorden, 2014). A Columbia university professor in biology and chemistry describes how they used Twitter to be alerted about the literature:

> Sometimes four or five people I follow will mention a paper that I did not come across and I will look it up. I think I am much more up to date on science literature since I started following Twitter. (Bonetta, 2009, p. 452)

A study by Tenopir, Volentine, and King (2013) found that academics on Twitter read more scholarly publications, which seems to confirm the use of Twitter as a publication alert service. At the same time, as the analyses by Priem and Costello (2010) and Letierce et al. (2010) show, the share of tweets linking to academic papers is low, suggesting a rather passive use: scholars follow links to tweeted articles but do only infrequently distribute them themselves. This suggests that a significant part of Twitter use cannot be captured by scholarly Twitter metrics.

#### *1.2.2.2 Retweets*

Retweets represent a specific form of diffusing information, as users forward messages sent by others. As such, they do not represent an original contribution by the retweeting user. Since retweets directly quote another users text, they can be seen as “internal citations” (Weller & Puschmann, 2011, p. 3) on Twitter. An analysis of retweets demonstrates how information circulates within a specific user community (Paßmann, Boeschoten, & Schäfer, 2014). A common disclaimer that *retweets do not equal endorsements* adapted from early Twitter use by journalists, emphasizes that tweets are forwarded to increase information diffusion. Once a frequent part of Twitter bios, the disclaimer has now been established as common sense and is no longer needed (Warzel, 2014).

As sharing information is one of the main motivations for scholarly Twitter use, retweeting is likely to be common among tweeting academics. Conference participants interviewed by Letierce and colleagues retweeted “tweets that are close to their interest or tweets that speak about their own work or research project” (Letierce et al., 2010, p. 7). Studies showed that retweeting is less common than other affordances used by scholars on Twitter but exceed expectations of a random sample of tweets in 2009, in which as few as 3% were retweets (boyd et al., 2010). Between 15% and 20% of tweets sent at scientific conferences were retweets (Letierce et al., 2010); similarly, 15% of 68,232 tweets by a group of 37 astrophysicists were retweets (Haustein, Bowman, Holmberg, et al., 2014). Retweeting was more common among 28 academics analyzed by Priem and Costello (2010), as 40% of tweets were retweets. Similarly, 37% of 43,176 tweets by Canadian doctoral students were retweets, while 10% of their tweets were retweeted (Work et al., 2015). At the same time, the share of retweets at a radiology conference was 60% (Hawkins et al., 2014). Being asked about Twitter affordance use, 14% of US professors said that they mostly or always and 34% that they sometimes retweeted (Bowman, 2015b). These professors were more than twice as likely to retweet when their tweets were classified as professional rather than personal (Bowman, 2015b). These differences again demonstrate that tweeting behavior differs depending on who is tweeting and it what context.

The majority of retweets sent by scholars from ten disciplines contained links, while conversational tweets (i.e., @mentions) were less likely to contain links.

> This clearly shows that researchers […] frequently share web content and forward information and content they have received from people they follow on Twitter, while links are not that often shared in conversations. (Holmberg & Thelwall, 2014, p. 1035)

Links to papers were significantly less likely to be retweeted: while 19% of tweets with links to scholarly publications were retweeted, the retweet rate was twice as high in the overall sample of tweets analyzed by Priem and Costello (2010). This is in contrast to a random sample of more than 200,000 tweets, over half of which contained a URL (boyd et al., 2010). A random sample of 270 tweets linking to scientific journal articles found that many were modified retweets of tweets originating from the journal's own Twitter account (Thelwall et al., 2013).

From the perspective of scholarly metrics, a distinction should arguably be made between tweets and retweets, as the latter reflects a rather passive act of information sharing (Holmberg, 2014). Although with each retweet the visibility of the tweet and the information it contains (e.g., the link to a publication) increases, retweets represent diffusion of information rather than impact. As retweeting requires even less effort—as little as one click since the implementation of the retweet button—than composing an original tweet, Twitter metrics should distinguish between tweets and retweets to reflect these different levels of user engagement (Haustein, Bowman, & Costas, 2016).

#### *1.2.2.3 Followers, @mentions and @replies*

Scholars on Twitter actively seek new connections and connect others (Veletsianos, 2012). Twitter is built in a way that information is spread via user networks. This means that users receive updates from those they follow and diffuse their messages to those they are followed by, creating a personal public (Schmidt, 2014). Selecting who to follow and who one is followed by are thus essential to communicating on Twitter. Twitter users build a reputation based on both their number of followers and followees. A parallel can be drawn to authors' citation identity and citation image based on which authors one cites and is cited by (White, 2000, 2001).

The purpose of expanding one's social network and finding peers was apparent in a study of 632 emergency physicians on Twitter: those who included work-related information in their Twitter bios had more followers and the most influential users in the network were connected to at least 50 other emergency physicians (Lulic & Kovic, 2013). Similarly, conference participants on Twitter saw a significant increase in their number of followers, particularly when they were also presenting (Sopan et al., 2012). The average number of followers of 260 physicians analyzed by Chretien et al. (2011) was 17,217 with a median of 1,426, indicating the typically skewed distribution of followers among Twitter users.

In addition to broadcasting one's message via retweets on a meso-level of communication, Twitter users can also directly address users with @mentions or @replies on an interpersonal level. Both replies and mentions thus represent a particular type of tweet that focuses on conversation rather than broadcasting (Honeycutt & Herring, 2009). Like most Twitter affordances, these types of tweets were also developed by the user base before being implemented by Twitter. While @replies happen in response to a tweet and are only visible on the timeline of the tweeter who sent the original tweet, @mentions refer to tweets that contains another user's Twitter handle, which triggers a notification to inform them about being mentioned (Bruns & Moe, 2014). About one third of tweets in 2009 included another user name (boyd et al., 2010). A random sample of tweets without an @mention were mostly about the tweeting user's experience, while those with an @mention were more likely about the addressee. In fact, more than 90% of @mentions functioned to address another user, while 5% worked as a reference (Honeycutt & Herring, 2009).

The great majority of tweets sent by social sciences and humanities doctoral students referred to other Twitter users, as 72% of the 43,176 tweets sent contained other user names (Work et al., 2015). Conversational tweets were also popular among a group of tweeting astrophysicists. Of the 68,232 tweets 46% were @replies or @mentions (61% including RTs), making it the most frequently used Twitter affordance. Conversational tweets were particularly common among those who tweeted regularly or frequently (Haustein, Bowman, Holmberg, et al., 2014). Most of these mentions referred to science communicators (24%), other astrophysicists (22%) or organizations (13%) on Twitter (Holmberg et al., 2014).

Conversational tweets hardly contained links (Holmberg & Thelwall, 2014), and among tweets linking to publications, only 8% were @replies (Priem & Costello, 2010). Opposed to hashtag use, retweeting and embedding URLs, @mentions were the only affordance that were less likely to occur in professors' professional tweets (56%) than those identified as personal (67%), which suggests that when professors discuss their work, they are less likely to address or reference other users directly than when they tweet about private matters. However, mentioning other users was still more common in their professional tweets than retweeting and using hashtags (Bowman, 2015b).

#### *1.2.2.4 Hashtags*

Similar to retweets and @mentions, hashtags are a user-driven Twitter affordance. Hashtags are keywords following the #-sign, which facilitate connections between users interested in the same topics. Conversations revolving around hashtags represent the macro layer of Twitter communication (Bruns & Moe, 2014). Holmberg and colleagues suggest that "hashtags may resemble the traditional function of metadata by enhancing the description and retrievability of documents" (Holmberg et al., 2014, p. 3).

Hashtag use seem to be less common than that of other Twitter affordances among academics. Sixty-one percent of surveyed American university professors declared that they rarely or never used a hashtag (Bowman, 2015b). This might be because scholars are either less familiar with this Twitter-specific affordance or they do not wish to expand conversations beyond their personal publics defined by their follower networks. Although actual hashtag use was low by the US professors analyzed by Bowman (2015b), they were more likely to use hashtags in their tweets identified as professional (28%) than those coded as personal (17%).

An early large-scale study found that among a random sample of 720,000 tweets, only 5% contained a hashtag (boyd et al., 2010). Almost one quarter of tweets by astrophysicists contained a hashtag (Haustein, Bowman, Holmberg, et al., 2014), but hashtag use varied significantly among different clusters of the follower network (Holmberg et al., 2014). The same share of hashtags (25%) was found for tweets of SSHRC funded doctoral students; on average each hashtag was mentioned 2.8 times (Work et al., 2015). It is problematic to infer hashtag use from most other studies on scholarly tweets, as data collection itself is often based on a specific hashtag, such as a conference hashtag (e.g., Letierce et al., 2010; Weller et al., 2011).

### 1.2.3 *Reluctance against and negative consequences of using Twitter*

When using tweets as the basis to measure scholarly impact of any sorts, it is essential to consider who is not on Twitter and why academics might be reluctant to join the microblogging platform. Twitter is a platform where "content is not king" (V. Miller, 2008, p. 395) and has been perceived as "shallow media, in the sense that it favors the present, popular and the ephemeral" (Rogers, 2014, p. xiv). Early Twitter studies which identified the majority of tweets to be "pointless babble" (Kelly, 2009, p. 5) or "daily chatter" (Java, Song, Finin, & Tseng, 2007, p. 62) casted doubt on the value of Twitter as a meaningful communication medium (Rogers, 2014). This reduction to banal content has led many in academia to consider tweeting a waste of time and therefore rejecting Twitter as a means of scholarly communication (Gerber, 2012). Particularly the 140-character limit of tweets has many scholars doubt Twitter's usefulness for research. This may be why Twitter is one of the best known and at the same time least used social media platforms in the scholarly community (Carpenter et al., 2012; Gerber, 2012; Gu & Widén-Wulff, 2011; Pscheida et al., 2013; Van Noorden, 2014).

The adoption of new technologies is often met with functional and psychological barriers. Lack of time and skills as well as a negative perception of platforms have been identified as barriers to use social media in academia (Donelan, 2016). Reluctance often stems from the notion that tweeting wastes precious time and introduces challenges that come with the blurred boundaries of professional and personal communication. This mix of professional and personal identities on Twitter specifically has been revealed by many studies and has been identified as a potential reason not to use Twitter in academia. Even though in the general public, Twitter uptake is higher among young adults, it is often early career researchers who are more reluctant to tweet about their work (Bulger et al., 2011; Coverdale, 2011). Young academics feel the highest pressure to publish in high-impact journals and limit their time spent on social media, and feel more vulnerable when publicly exposing their ideas, particularly to uncertain audiences (Coverdale, 2011; Harley, Acord, Earl-Novell, Lawrence, & King, 2010; Mccrea, 2011; Veletsianos, 2013). An academic interviewed about the future of scholarly communication expressed their concern regarding the use of Facebook and Twitter for work purposes:

> There's this research group in my area and, for some reason, they're really into Facebook. So they want to do a lot of discussions on Facebook and that type of thing. But I really just don't have time. It's like Twittering. I just can't…There needs to be a little bit of space where I can actually think about something. And I think for some people, they're just wired in such a way that they like that constancy, and they are also able to actually say something intelligent quickly. And I'm not like that. I have to be a little bit more deliberate and think about things a little bit more. And so I can't Twitter…I need some time to reason… (Harley et al., 2010, p. 97)

However, early-career researchers were often more likely to find social media useful in the context of scholarly communication and collaboration (Grande et al., 2014; Gruzd, Staves, & Wilk, 2012; Tenopir et al., 2013). Bowman (2015b) identified a u-shaped relationship between Twitter use and academic experience: US professors seven to nine years into their academic careers were more likely to use Twitter compared to those with fewer and more years of experience. A humanities scholar addressed the issue of blurred boundaries on Twitter:

> I think it can be distracting, especially to grad students, when they're trying to navigate, when they're needing to learn, adopt, and use these new technologies, but at the same time learn to discriminate among technologies that are more for social things but are being used in the name of research. The lines are too blurry. (Bulger et al., 2011, p. 59)

The tension that arises between scholars who take to Twitter and those who are reluctant to discuss scholarly matters on social media is demonstrated by an incident where a genomics paper published in *PNAS* was criticized on Twitter for flaws in study design and analysis casting doubt on its conclusions. The tweet sent by a genetics researcher at the University of Chicago included charts and tables from a re-analysis of the *PNAS* paper's data, which he published in the open access and open-peer review journal *F1000Research*. The tweet and re-analysis provoked many responses on Twitter and in the comment section of the *F1000Research* paper, some of which demanded the retraction of the *PNAS* paper (Woolston, 2015), while the *PNAS* authors accused the critic of violating the norms of science by taking to Twitter.

The clashing of personal and professional and the fuzzy boundaries between the two has also lead to severe negative consequences for scholars. Tweets by faculty have caused outrages among students, other faculty members, university administration and the public at large. Identifying professionally on Twitter has affected the academic careers of some scholars and, in a certain case, controversial tweets have provoked death threads (Rothschild & Unglesbee, 2013). Tweeting had serious effects on the career of a tenured University of New Mexico psychology professor, who had fat-shamed students on Twitter: "Dear obese PhD applicants: if you didn't have the willpower to stop eating carbs, you won't have the willpower to do a dissertation #truth." The professor was asked to apologize and had to undergo sensitivity training, while his work was monitored and he was banned from working on the graduate students admission committee for the rest of his career (Bennett-Smith, 2013; Ingeno, 2013; Pomerantz et al., 2015).

In another incident, an English professor lost his tenure-track position offer from the University of Illinois at Urbana-Champaign due to a tweet that was interpreted as anti-Semitic (Herman, 2014). In response to withdrawal of the job offer by the University of Illinois, the American Association University Professors updated their report on academic freedom to reinstate that academic freedom applies to comments made on social media (Association of American University Professors, 2013).

## 1.3 Scholarly output on Twitter

As this chapter discusses scholarly Twitter metrics, it focuses on how scientific papers are diffused via tweets. Although, as described above, only a small amount of scholars' tweeting activity involves linking to publications (Letierce et al., 2010; Priem & Costello, 2010), tweets to journal articles represents one of—if not *the*—most popular altmetric. This might be due to the significance of publications in peer-reviewed journals in scholarly communication as well as the ease at which tweets that mention or link to document identifiers (e.g., the Digital Object Identifier DOI) can be retrieved. Another reason that Twitter-based altmetrics—and, in fact, all altmetrics—gravitate towards journal articles is that they aim to complement existing bibliometric measures, which reduce scholarly output in a similar manner.

In the following, the altmetrics literature is reviewed to provide an overview of currently used scholarly Twitter metrics for journal articles. As the majority of available studies only scratch the surface of what can potentially be extracted from Twitter activity, the literature review is complemented by an analysis of tweets collected by Altmetric.com. This analysis goes beyond tweet counts and correlations with citations and aims to reflect the *What*, *How*, *When*, *Where* and *Who* of scholarly publications shared on Twitter. This includes what types of documents are tweeted, how Twitter affordances such as hashtags, retweets and @mentions, are used to share them, when and where articles are tweeted and who is tweeting them. Although this chapter is not able to provide solutions to problems associated with creating meaningful metrics from tweets, particular limitations and pitfalls are highlighted, while demonstrating the potential of available data. Together with the findings on Twitter use described above, the chapter tries to provide context to help the interpretation of different metrics. This chapter thus aims at improving the understanding of what Twitter-based scholarly metrics can and cannot reflect.

### 1.3.1 Data and indicators

The analysis of tweets that mention scholarly documents is based on Twitter data collected by Altmetric.com until June 2016, which contains 24.3 million tweets mentioning 3.9 unique documents. Altmetric started systematically collecting online mentions of scholarly publications in 2012 and is a particularly valuable data source for tracking Twitter activity related to scholarly output, as it continuously stores tweets that mention scholarly publications with a DOI. Through accessing tweets through the Twitter API firehose, Altmetric circumvents the usual issues of Twitter data collection that researchers are confronted with when using the freely available Twitter APIs. While the Twitter *Streaming* API limits access to a random sample of 1% of tweets, the *Representational State Transfer* (REST) API is rate-limited and the *Search* API restricts access to only the most recent tweets relevant to a particular query (Gaffney & Puschmann, 2014). As Altmetric started data collection in 2012, Twitter activity is incomplete for documents published earlier.

**Table 1** Scholarly Twitter metrics associated with scholarly documents

| **Type of metric** | **Scholarly Twitter metric** | **Description** |
|---|---|---|
| Tweets | Twitter coverage | Percentage of documents with at least one tweet |
| | Number of tweets | Sum of total number of tweets |
| | Twitter density | Mean number of tweets per document |
| | Twitter intensity | Mean number of tweets per tweeted document |
| | | |
| Retweets | Share of retweets | Percentage of tweets that were retweets |
| | Retweet density / intensity | Mean number of retweets per document / tweeted document |
| | | |
| Users | Number of users | Unique number of users associated with a document |
| | User density / intensity | Mean number of users per document / tweeted document |
| | Mean number of followers | Mean of the number of followers of users tweeting a document |
| | | |
| Hashtags | Hashtag coverage | Percentage of documents with at least one hashtag |
| | Number of hashtags | Unique number of hashtags associated with a document |
| | Hashtag frequency | Sum of total number of hashtag occurrences |
| | Share of hashtags | Percentage of tweets with at least one hashtag |
| | Hashtag density / intensity | Mean number of hashtags per document / tweeted document |
| | | |
| @mentions | @mention coverage | Percentage of documents mentioning a user name |
| | Number of mentioned users | Unique number of users mentioned in tweets associated with a document |
| | @mention frequency | Sum of total number of @mentions |
| | @mention density / intensity | Mean number of @mentions per document / tweeted document |
| | | |
| Time | Tweet span | Number of days between first and last tweet |
| | Tweet delay | Number of days between publication of a document and its first tweet |
| | Twitter half-life | Number of days until 50% of all tweets have appeared |

Altmetric's Twitter data is matched to bibliographic information from WoS using the DOI. This match between document metadata and tweets affords the possibility to determine the amount of scholarly output that does and does not get tweeted. The link to WoS data also provides access to cleaner and extended metadata of tweeted documents, such as the publication year, journal, authors and their affiliations, and a classification system of scientific disciplines. At the same time, the match of the two databases also excludes tweets to publications not indexed in WoS and thus comes with the known restriction and biases of WoS coverage. This is why the following analysis describes results for two datasets, containing, respectively all 24.3 million tweets covered by Altmetric (dataset A), and 3.9 million tweets mentioning documents with a DOI, covered by WoS and published in 2015 (dataset B). The number of unique documents in dataset A is based on Altmetric ID. As Altmetric.com's metadata is based on multiple sources, one publication might be treated as two documents, particularly if it has multiple versions, such as a journal article on the publisher's website and a preprint on *arXiv*. Similar duplications are possible but less likely in dataset B, which is based on unique identifiers and cleaner metadata in WoS.

In the following, a set of descriptive indicators are used based on tweets to scholarly documents and associated metadata. Table 1 provides an overview of each metric. As described above, the focus here is on journal articles but the metrics can, nevertheless, be applied to any scholarly document or other research object such as scholarly agents including "individual scholars, research groups, departments, universities, funding organizations and others entities acting within the scholarly community" (Haustein, Bowman, & Costas, 2016, p. 376). Similar to most bibliometrics, the metrics described in Table 1 can be applied to any aggregated set of documents, such as all documents relevant to a certain topic, published in the same journal, by the same author, institution, country or in a specific language.

### 1.3.2 What scholarly output is tweeted?

As shown above, only a small share of tweets sent by scholars actually link to scholarly output. Tweets linking to blogs or other websites are often more frequent than those linking to scientific publications. At the same time, the great majority of altmetric studies on Twitter focus on peer-reviewed scientific journal articles with a DOI, which represent one of the main limitations of currently captured altmetrics (Alperin, 2013; Haustein, 2016; Haustein, Sugimoto, & Larivière, 2015; Taylor, 2013).

In comparison to other common altmetrics, Twitter is the platform which exhibits the second largest activity related to scientific papers, following the social reference management platform Mendeley (Costas, Zahedi, & Wouters, 2015b; Haustein, Costas, et al., 2015). The disciplinary differences in Twitter uptake described above are equally visible, and likely resulting in, large difference in Twitter coverage between scientific disciplines. In the majority of studies, usually between 10% to 30% of selected documents were mentioned on Twitter at least once (Alhoori & Furuta, 2014; Andersen & Haustein, 2015; Costas et al., 2015a; Hammarfelt, 2014; Haustein, Costas, et al., 2015; Haustein, Larivière, Thelwall, et al., 2014; Haustein, Peters, Sugimoto, et al., 2014; Priem, Piwowar, et al., 2012). As the majority of tweets linking to scholarly papers occurs immediately after publication (Eysenbach, 2011; Shuai et al., 2012) and Twitter activity increased annually, Twitter coverage increases by year of publication, with the most recent papers being more likely to be tweeted and older papers hardly getting shared on Twitter (Haustein, Peters, Sugimoto, et al., 2014). For example, while more than half of 2015 PLOS papers were tweeted at least once (Barthel et al., 2015), Twitter coverage was at 12% for those published in 2012 (Priem, Piwowar, et al., 2012). Similarly, 13% of 2011(Costas et al., 2015a), 16% of 2011-2013 (Robinson-García, Torres-Salinas, Zahedi, & Costas, 2014) and 22% of 2012 WoS documents (Haustein, Costas, et al., 2015) had received at least one tweet, while Twitter coverage increased to 36% for WoS 2015 papers in dataset B (see Table 2 below).

Coverage varied between disciplines and journals, but also between databases and geographic regions. Twitter is blocked in countries like Iran and China, which reflects on the visibility of their authors and papers. For example, the share of tweeted papers was low for papers published by authors from Iran (Maleki, 2014). Such geographical biases affect Twitter visibility and need to be taken into account when comparing Twitter impact of documents, authors and institutions from different countries. For example, as few as 6% of Brazilian documents published 2013 and indexed in SciELO (Alperin, 2015a) and 2% of a sample of Iranian papers covered by WoS had been tweeted (Maleki, 2014). On the contrary, at 21% Twitter coverage of Swedish publications seems to be more in line with general findings (Hammarfelt, 2014). As the sample of Iranian publication included documents published between 1997 and 2012 and 98% of tweeted papers were published between 2010 and 2012, both geographical and publication date biases influence Twitter coverage.

Particularly high coverage was found for *arXiv* submissions (Haustein, Bowman, Macaluso, Sugimoto, & Larivière, 2014; Shuai et al., 2012). This high activity was, however, not caused by high Twitter uptake in the physics, mathematics and computer science communities, but created by Twitter bots (Haustein, Bowman, Holmberg, et al., 2016). The extend of such automated rather than human activity and its implications for the meaning of scholarly Twitter metrics are further discussed below.

While Twitter coverage describes the extent to which a set of documents gets diffused on Twitter, Twitter density (i.e., the mean number of tweets per paper) reflects the average tweeting activity per document. In general, each paper receives less than one tweet on average, with large variations between disciplines (see below). As Twitter density is influenced by Twitter coverage and thus particularly low when only a small share of papers gets tweeted, Twitter intensity reflects the average tweeting activity for tweeted papers only, excluding non-tweeted papers (Haustein, Costas, et al., 2015). The number of tweets per paper is usually heavily skewed, much more than citations. For example, 63% of tweeted biomedical papers were only tweeted once (Haustein, Peters, Sugimoto, et al., 2014).

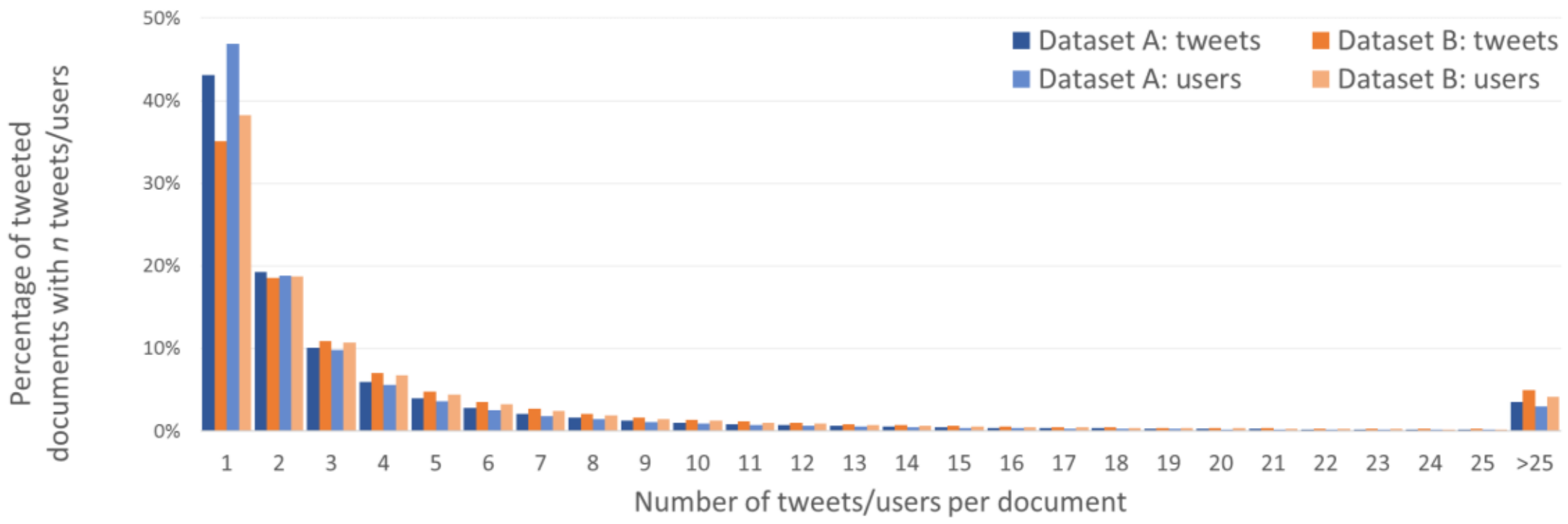

**Figure 1** Number of tweets and users per tweeted document for datasets A and B.

A similar distribution can be observed for the two datasets described above (Figure 1). The 24.3 million tweets in dataset A mentioned 3.9 documents (based on Altmetric ID), 43% of which were tweeted once, 19% twice, 10% three times, while only 4% of documents received more than 25 tweets. The document with the most Twitter activity was mentioned 35,135 times by 144 users, with one user tweeting 34,797 times. This highlights that the number of distinct users per document might be a better proxy of diffusion than the total number of tweets.

Tweets in dataset A were sent by 2.6 million users and link to 3.9 million documents, amounting to a Twitter intensity of 6.2 tweets per tweeted paper. The number of users is based on unique Twitter handles included as *Author ID on Source* in Altmetric's Twitter data. Since the data does not include Twitter's unique user ID, users might be counted more than once if they changed their Twitter handle. Very likely caused by the 140-character limitation, less than 1% of tweets link to more than one document. As shown in Figure 1, the distribution of number of users per document is even more skewed with 47% of all documents tweeted by a single user only. The distribution for WoS 2015 papers (dataset B) is similar but slightly less skewed for documents (based on WoS identifier) mentioned in one tweet and by one user only; 35% and 38% of documents were tweeted once or by one user, respectively. The total number of 3.9 million tweets were sent by 601,290 users mentioning 548,841 documents, which corresponds to a Twitter coverage of 36%, intensity of 7.2 and a density of 2.6 tweets per document.

### *1.3.2.1 Disciplines and journals*

Twitter activity varies among disciplines and even journals of the same field. As there is no gatekeeping in Twitter, these variations might not be entirely due to actual impact of a particular journal but can be heavily influenced by individuals and marketing strategies of publishing houses or other stakeholders. Despite these variations, studies have shown that multidisciplinary and biomedical journals as well as social science publications are particularly visible on Twitter, while the so-called hard sciences are tweeted about less (Alhoori & Furuta, 2014; Andersen & Haustein, 2015; Barthel et al., 2015; Costas et al., 2015b; Fenner, 2013; Hammarfelt, 2014; Haustein, Costas, et al., 2015;

Haustein, Larivière, Thelwall, Amyot, & Peters, 2014; Haustein, Peters, Sugimoto, et al., 2014; Knight, 2014; Priem, Piwowar, et al., 2012).

Previous findings are corroborated by the analysis of discipline and journal-based tweeting activity in dataset B. As shown in Table 2, over one third of 2015 documents in WoS have been tweeted, which represents a significant increase compared to papers published in the previous years. Twitter coverage shows large variation between disciplines from 59% of publications in Biomedical Research, Health and Psychology to less than 10% in Mathematics and Engineering & Technology. This upholds previous findings that Twitter activity is particularly elevated around publications from the biomedical and social sciences (Costas et al., 2015a; Haustein, Costas, et al., 2015).

**Table 2** Dataset B: Twitter activity for WoS articles published in 2015.

| | | Papers 2015 | | Tweeted papers | | Tweets | | | | Users | |
|---|---|---|---|---|---|---|---|---|---|---|---|
| **Discipline** | | N | DOI coverage | N | Twitter coverage | N | Density | Intensity | % RTs | N | Intensity |
| All disciplines | | 2,014,977 | 76% | 548,841 | 36% | 3,960,431 | 2.6 | 7.2 | 50% | 601,290 | 6.6 |
| Natural Sciences & Engineering | Biology | 124,402 | 73% | 33,945 | 37% | 226,575 | 2.5 | 6.7 | 54% | 52,235 | 4.3 |
| | Biomedical Res. | 226,011 | 84% | 112,470 | 59% | 1,025,061 | 5.4 | 9.1 | 50% | 229,851 | 4.5 |
| | Chemistry | 158,929 | 90% | 32,829 | 23% | 81,739 | 0.6 | 2.5 | 34% | 13,860 | 5.9 |
| | Clinical Medicine | 663,481 | 64% | 223,641 | 52% | 1,784,438 | 4.2 | 8.0 | 51% | 288,226 | 6.2 |
| | Earth & Space | 96,792 | 91% | 25,616 | 29% | 136,732 | 1.6 | 5.3 | 51% | 42,641 | 3.2 |
| | Engr. & Tech. | 253,020 | 88% | 18,441 | 8% | 50,577 | 0.2 | 2.7 | 33% | 17,235 | 2.9 |
| | Mathematics | 51,240 | 85% | 2,890 | 7% | 17,441 | 0.4 | 6.0 | 43% | 7,826 | 2.2 |
| | Physics | 128,766 | 93% | 16,783 | 14% | 55,295 | 0.5 | 3.3 | 30% | 14,822 | 3.7 |
| Social Sciences & Humanities | Arts | 18,995 | 19% | 566 | 15% | 1,962 | 0.5 | 3.5 | 39% | 1,134 | 1.7 |
| | Health | 51,535 | 74% | 22,662 | 59% | 191,530 | 5.0 | 8.5 | 52% | 60,090 | 3.2 |
| | Humanities | 76,998 | 39% | 4,601 | 15% | 19,150 | 0.6 | 4.2 | 52% | 9,784 | 2.0 |
| | Prof. Fields | 54,109 | 72% | 14,760 | 38% | 93,936 | 2.4 | 6.4 | 47% | 41,449 | 2.3 |
| | Psychology | 40,657 | 78% | 18,735 | 59% | 145,767 | 4.6 | 7.8 | 51% | 50,974 | 2.9 |
| | Social Sciences | 70,042 | 76% | 20,902 | 39% | 135,193 | 2.5 | 6.5 | 53% | 51,288 | 2.6 |

It should be emphasized that Twitter activity in the Arts and Humanities cannot be generalized as DOIs are not commonly used in these disciplines. While in general 76% of all documents had a DOI, as few as 19% of all WoS-indexed journal articles in the Arts were linked to this unique identifier. DOI use does not only differ between disciplines but also by country or language of publication, which is why results may be biased in these regards as wells. Considering *National Science Foundation* (NSF) specialties with more than half of its papers having a DOI, Twitter coverage was highest in Parasitology (78%), Allergy (76%) and Tropical Medicine (70%) and lowest in Metals & Metallurgy (1%), Miscellaneous Mathematics (2%), Mechanical Engineering (2%) and General Mathematics (3%). With more than 10 tweets per document, Twitter density was highest in General & Internal Medicine (13.5) and Miscellaneous Clinical Medicine (12.3).

At a Twitter coverage of 100%, 198 of 9,340 tweeted journals had all of their documents diffused on Twitter, which strongly suggests a systematic and automated diffusion, possibly by a dedicated account maintained by the journal or publisher (see below). Among journals with at least 100 papers with a DOI in 2015, *JAMA*, and *Biotechnology Advances* had the highest Twitter density at 115.4 and 113.2 tweets per paper, respectively. With more than 10,000 unique Twitter users, *PLOS ONE*, *BMJ*, *Nature*, *Science*, *PNAS*, *NEMJ*, *JAMA*, *Lancet*, *Scientific Report*, *Nature Communications*, *JAMA Internal Medicine*, *PLOS Biology*, *British Journal of Sports Medicine*, *Cell*, *Biotechnology Advances* and *BMJ Open* were tweeted by the largest audience on Twitter (Table 3). These journals reflect large multidisciplinary and biomedical journals with large readership, which have been previously identified as highly tweeted, possibly because they exhibit a particular relevance to people's everyday lives (Costas et al., 2015b). Almost all of these journals maintain official Twitter accounts appearing among the three most tweeting users, which suggests that Twitter has become an important platform for journals and publishers to promote their contents. The number of unique Twitter users, in particular for the general science journals seems to be low in comparison to global readership. For example, while *Nature* claimed to have 3 million unique visitors per month (Nature Publishing Group, 2012), as few as 42,365 Twitter users mentioned a 2015 paper. However, numbers seem to be in line with print circulation (Nature Publishing Group, 2010).

**Table 3** Twitter metrics for journals with the largest Twitter audience based on number of users (≥10,000 users) including three most active users. Journal accounts are highlighted in bold.

| Journal | Twitter coverage | Tweets | | | | Users | |
|---|---|---|---|---|---|---|---|
| | | N | Density | Intensity | %RTs | N | Top 3 most active users (according to number of tweets; official journal accounts are marked in bold) |
| PLOS ONE | 70% | 148,494 | 5.2 | 7.4 | 47% | 59,210 | aprendedorweb, uranus_2, **PLOSONE** |
| BMJ | 91% | 149,874 | 45.3 | 49.7 | 66% | 57,622 | **bmj_latest**, bookapharmacist, npceorg |
| Nature | 100% | 85,523 | 89.0 | 89.3 | 56% | 42,365 | Doyle_Media, randomshandom, TheRichardDoyle |
| Science | 99% | 78,141 | 70.0 | 70.8 | 67% | 39,225 | **sciencemagazine**, PedroArtino, rkeyserling |
| PNAS | 89% | 72,312 | 19.9 | 22.4 | 54% | 36,496 | abbrabot, EcoEvoJournals, uranus_2 |
| NEMJ | 99% | 74,263 | 75.2 | 75.9 | 62% | 33,142 | medicineupdate, **NEJM**, JebSource |
| JAMA | 95% | 68,420 | 115.4 | 121.1 | 64% | 31,018 | **JAMA_current**, robarobberlover, ehlJAMA |
| Lancet | 98% | 45,752 | 69.9 | 71.6 | 65% | 25,673 | medicineupdate, darmtag, **TheLancet** |
| Scientific Reports | 56% | 41,177 | 4.6 | 8.3 | 50% | 23,395 | **SciReports**, uranus_2, geomatlab |
| Nature communications | 78% | 39,540 | 11.8 | 15.1 | 52% | 21,274 | **NatureComms**, Kochi_Study, PatrickGoymer |
| JAMA Internal Medicine | 91% | 48,882 | 99.8 | 110.1 | 69% | 20,197 | **JAMAInternalMed**, JAMA_current, GeriatriaINNSZ |
| PLOS Biology | 99% | 24,481 | 90.0 | 91.3 | 70% | 12,878 | **PLOSBiology**, PLOS, nebiogroup |
| BJ of Sports Medicine | 97% | 26,121 | 75.7 | 77.7 | 69% | 11,618 | **BJSM_BMJ**, exerciseworks, SportScienceNI |
| Cell | 96% | 23,519 | 38.2 | 39.9 | 57% | 11,081 | Brianxbio, **CellCellPress**, topbiopapers |
| Biotechnology Advances | 55% | 14,823 | 113.2 | 205.9 | 67% | 10,331 | robinsnewswire, GrowKudos, ElsevierBiotech |
| BMJ Open | 76% | 16,882 | 11.5 | 15.1 | 60% | 10,090 | **BMJ_Open**, LS_Medical, SCPHRP |

Going beyond peer-reviewed journals indexed in WoS, other sources (based on Altmetric metadata) of scholarly documents are also frequently shared on Twitter. In fact, the largest number of tweeted documents in dataset A came from *arXiv*: a total of 319,411 *arXiv* submissions were tweeted 1.1 million times by 110,134 users. As shown in Figure 2, the number of tweeted documents and unique users derivate particularly for the most popular sources. Although *arXiv*, *PLOS ONE* and *SSRN* are the most popular platforms according to the number of tweeted documents, *Nature*, *The Conversation* and *PLOS ONE* are tweeted by the largest number of users.

The majority of popular sources are peer-reviewed journals indexed in WoS, which also lead the ranking in dataset B (see above). Apart from these, links to repositories for documents (arXiv, SSRN, bioRxiv) or data (figshare, Dryad) as well as to website like *The Conversation* or *ClinicalTrials.gov* are also frequently tweeted. It should be mentioned that the document metadata in Altmetric is based on a variety of sources and thus shows some inconsistencies. For example, the source is unknown for 6% of tweeted documents and 4% of tweets in dataset A, and some sources appear in various spellings. Therefore, the number of distinct sources (49,379) or journal IDs (28,457) represents an overestimation of different sources. These inconsistencies can also be found in the publication year and other document metadata, which are essential to characterize what kind of scholarly output has been diffused on Twitter.

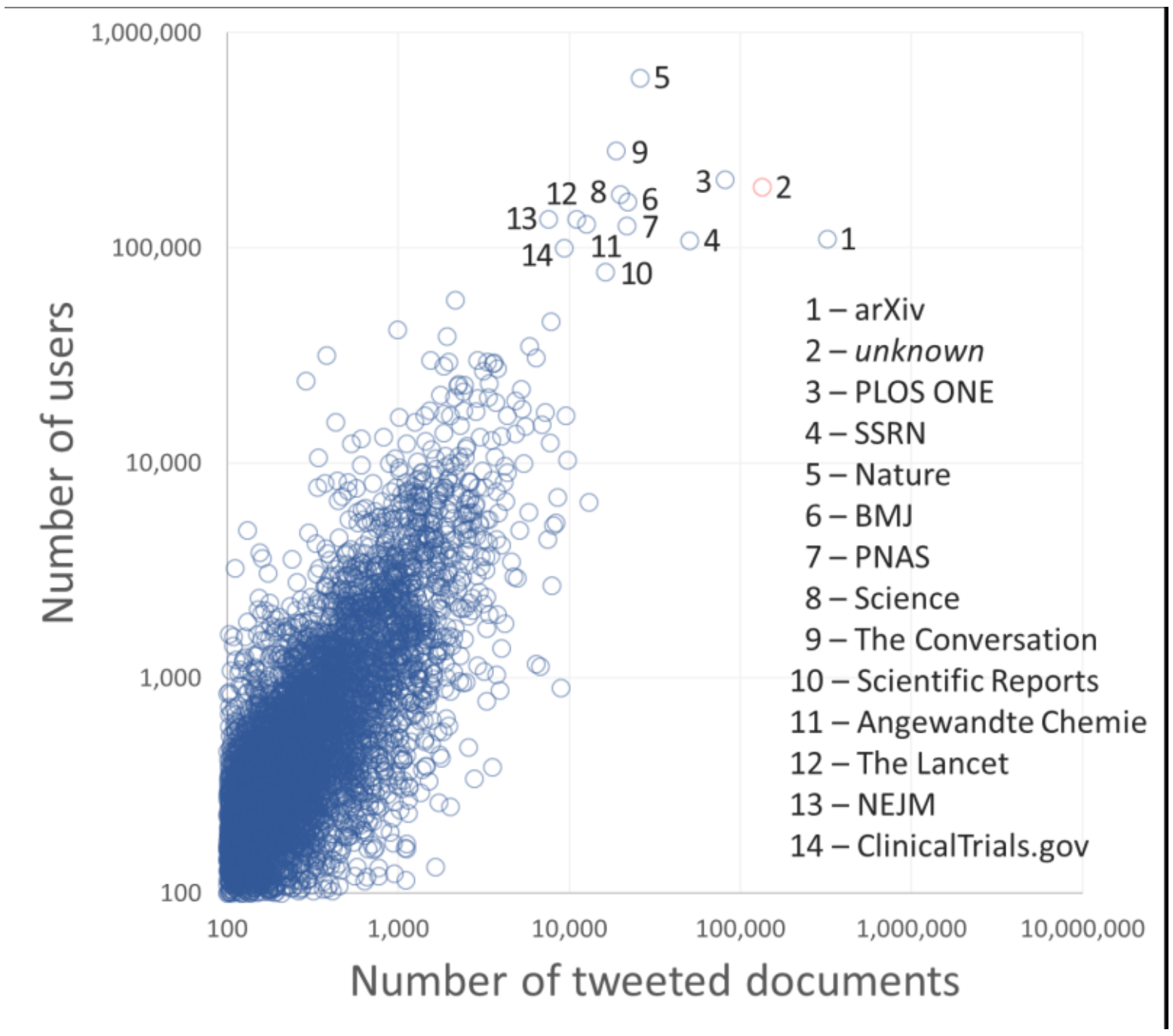

**Figure 2** Number of tweeted documents and unique users per source in dataset A (≥100 tweeted documents).

### *1.3.2.2 Document characteristics*

Besides disciplinary and topical differences, studies have also focused on determining what type of documents are popular on Twitter. Papers with particular high Twitter visibility often had humorous or entertaining contents rather than scientific merit (Haustein, Peters, Sugimoto, et al., 2014; Neylon, 2014). A study that coded title characteristics of 200 highly tweeted papers found that 16 included a cultural reference (i.e., proverbs, idioms, fictional characters, music) and 13 were humorous or light (Didegah, Bowman, Bowman, & Hartley, 2016). Bornmann (2014) reported that among papers recommended on F1000, those labeled as *good for teaching* were frequently tweeted. Andersen and Haustein (2015) found that meta-analysis and systematic reviews received significantly more tweets than other medical study types.

Marking a clear distinction from citation patterns, particularly high Twitter activity was found for document types that are usually considered *uncitable* (Haustein, Costas, et al., 2015). For example, Twitter coverage for news items was twice as high as that for all documents. Twitter density was highest for news items (3.0), editorial material (1.6) and reviews (1.4), by far exceeding the overall average of 0.8 tweets per document. The success of these document types on Twitter suggests that "documents that focus on topical subjects, debates and opinions, which are probably presented in simpler and less technical language, are more likely to appear and become popular on Twitter" (Haustein, Costas, et al., 2015, p. 8). Publications with shorter titles, fewer pages and references tend to receive more tweets, while the opposite tendencies are usually observed for citations (Haustein, Costas, et al., 2015).

### *1.3.2.3 Similarity to other types of usage*

Since altmetrics were proposed as an alternative or complementary to traditional bibliometric indicators, most studies analyze in how far these new impact metrics correlate with citations. The motivation behind correlation studies lies in determining whether tweeting patterns are comparable to citing behavior. Positive correlations between citation and tweets would indicate that tweets measure something similar to but much earlier than citations, making tweet-based indicators predictors of impact on the scholarly community (Eysenbach, 2011). Early studies argued that low or negative correlation coefficients would indicate a different type of impact than that on citing authors, possibly impact on society in general.

The first study analyzing the relationship between tweets and citations found a significant association between highly tweeted and highly cited papers published in the *Journal of Medical Internet Research*, as highly tweeted publications identified 75% of those which were later highly cited (Eysenbach, 2011). However, this claim was based on 55 papers in a journal which itself maintained a strong Twitter presence. For a set of 4,606 *arXiv* submissions, tweets were a better predictor of early citations than downloads (Shuai et al., 2012). The generalizability of the finding that tweets were early indicators of citation impact was later refuted by a large-scale analysis based on 1.3 million documents, which found overall low correlations between tweets and citations (Haustein, Peters, Sugimoto, et al., 2014).

Correlations between tweets and citations vary between datasets due to particular differences between disciplines or journals but are overall low between 0.1 and 0.2 (Barthel et al., 2015; Costas, Zahedi, & Wouters, 2015a; Haustein, Costas, et al., 2015a; Haustein, Peters, Sugimoto, et al., 2014; Priem, Piwowar, et al., 2012). It should be noted that correlations are affected by low Twitter coverage and thus differ whether untweeted papers are excluded or included from the analysis (Haustein, Costas, et al., 2015). Instead of replacing citations as a faster and better filter of relevant publications, as was suggested in the altmetrics manifesto (Priem, Taraborelli, Groth, & Neylon, 2010), the number of tweets seem to mirror visibility on other social media platforms, in particular Facebook, rather than visibility within the community of citing authors (Barthel et al., 2015; Haustein, Costas, et al., 2015a). If nothing else, the difference between tweet and citation counts as reflected in low correlations might be due to the fundamental difference between the act of citing and tweeting (Haustein, Bowman, & Costas, 2016).

A moderate negative correlation was found comparing publication output and tweeting activity of a group of astrophysicists on Twitter, suggesting that researchers who tweet a lot focus their efforts on communication and outreach rather than publishing peer-reviewed articles (Haustein, Bowman, Holmberg, et al., 2014). This inverse relationship between a researchers' standing in the scholarly community and their visibility on Twitter has led to the so-called Kardashian index, a tongue-in-cheek indicator that reveals that those who tweet more publish less and vice versa (Hall, 2014).

Rather than correlating tweets and citations Allen, Stanton, Di Pietro and Mosley (2013) aimed to measure the effect of promoting articles on social media (including Twitter) on usage statistics. Comparing the number of views, downloads and citations of randomly selected articles published in PLOS ONE before and after promoting them on social media, article views and downloads increased significantly but citations one year after publication and social media metrics did not.

### 1.3.3 How is scholarly output tweeted?

Not the least due to the evaluation community's focus on counts, altmetrics research has focused much less on tweet content than on correlations and other quantitative measures. Among those looking at tweet content, the focus has been on the analysis of Twitter specific affordance use (Bowman, 2015b; Weller et al., 2011). This includes in particular the use and analysis of hashtags, retweets and @mentions, which are further described below.

Analyzing 270 tweets linking to journal articles, Thelwall et al. (2013) found that 42% contained the title of the article, 41% summarized it briefly and 7% mentioned the author. As few as 5% explicitly expressed interest in the article. While sentiment was absent in the great majority of tweets, 4% of tweets were positive and none negative. Similarly a large-scale study that automatically identified sentiments in tweets using *SentiStrength* found that the majority of tweets were neutral and that, if sentiment was expressed, it was positive rather than negative (Friedrich, 2015). Based on 192,832 tweeted WoS documents published in 2012, 11% of 487,610 tweets were positive, 7% negative and 82% did not express any sentiment after removing the article title words from tweets. Tweets linking to chemistry papers were the least likely to express sentiments (Friedrich, 2015; Friedrich, Bowman, & Haustein, 2015).

### *1.3.3.1 Retweets and @mentions*

As described above, retweets and @mentions represent a particular form of conversational tweets, which seemed to enjoy particular popularity among academic Twitter use. Half of the 4 million tweets linking to 2015 WoS papers were retweets (Table 2), which suggests that a significant amount of tweeting activity reflects information diffusion that does not involve much engagement. Compared to the studies investigating retweet use among general Twitter users (boyd et al., 2010) and academics (Haustein, Bowman, Holmberg, et al., 2014; Letierce et al., 2010; Priem & Costello, 2010; Work et al., 2015), the share of retweets among tweeted journal articles is rather high. The percentage of retweets tends to be lowest in disciplines with low Twitter coverage, which suggests that users in disciplines with low Twitter uptake do not use it as much for information diffusion, possibly because they are not as well connected. In 32 of 120 NSF specialties with DOI coverage above 50%, retweets exceed original tweets (Table 2): retweeting was particularly common in Miscellaneous Zoology, General & Internal Medicine, Miscellaneous Clinical Medicine and Ecology with retweet rates above 60% and low in Solid State Physics, Inorganic & Nuclear Chemistry, Chemical Physics and Applied Chemistry with less than 20%.

### *1.3.3.2 Hashtags*

Thirty-one percent of the 24.3 million tweets captured by Altmetric until June 2016 contained a hashtag, which is comparable to other studies on hashtag use by academics (Haustein, Bowman, Holmberg, et al., 2014; Work et al., 2015) but far higher than the 5% among a random sample of tweets in 2009 (boyd et al., 2010). 401,287 unique hashtags were mentioned 12.6 million times, which amounts to an average occurrence of 31 per unique term. 105,705 unique hashtags were used in tweets linking to 2015 WoS papers. While 33% tweets contained at hashtag, 46% of all articles were described with at least one hashtag. Each hashtag was mentioned on average 21 times for a total hashtag frequency of 2.2 million. Hashtag frequency is extremely skewed, as 3% and 6% of hashtags are responsible for 80% of hashtag occurrences in dataset A and B, respectively. For example, the most popular hashtag in dataset A was used 162,754 times (1.3% of all occurrences), while 169,992 hashtags only occurred once. Figure 3 demonstrates on a log-log scale the number of tweets hashtags were mentioned in, as well as the number of distinct users mentioning each hashtag. While in general, a linear relationship can be found between the number of occurrences and users, a few popular hashtags are tweeted only by limited number of users, indicating a smaller community.

The most popular hashtags in dataset A were #science (Table 4), 1.3% of hashtag occurrence), #cancer (0.9%), #physics (0.8%), #openaccess, #health (0.7%), #paper, #oa and #research (0.5% each). The occurrence of #oa as well as #openaccess among the most frequent hashtags reflects the known heterogeneity of folksonomies and the need for *tag gardening* when trying to analyze topics (Peters, 2009). WoS 2015 papers (dataset B) were most frequently tagged as #cancer (1.0%), #health, #openaccess, #science (0.9%), #FOAMed, #Diabetes, #ornithology and #Psychiatry (0.6%). The order changes, when considering the number of unique users instead of tweets per hashtag. Among hashtags that occurred at least 1,000 times, the largest discrepancy between the number of tweets and users can be observed for #genomeregulation (1,924 tweet; 10 users), #eprompt (2,281; 17) and #cryptocurrency (4,515; 38), which were, on average, tweeted more than 100 times by the same users. On the contrary, the user-hashtag ratio was lowest for #Fit (4,818; 4,743), #StandWithPP (*Stand with Planned Parenthood*; 1,060; 972), #dataviz (1,010; 912), #coffee (1,517; 1,246), and #PWSYN (title of popular science book *The Patient Will See You Now*; 1,017; 834), which indicates a widespread adoption among Twitter users. Accordingly, these hashtags are more general and less scientific.

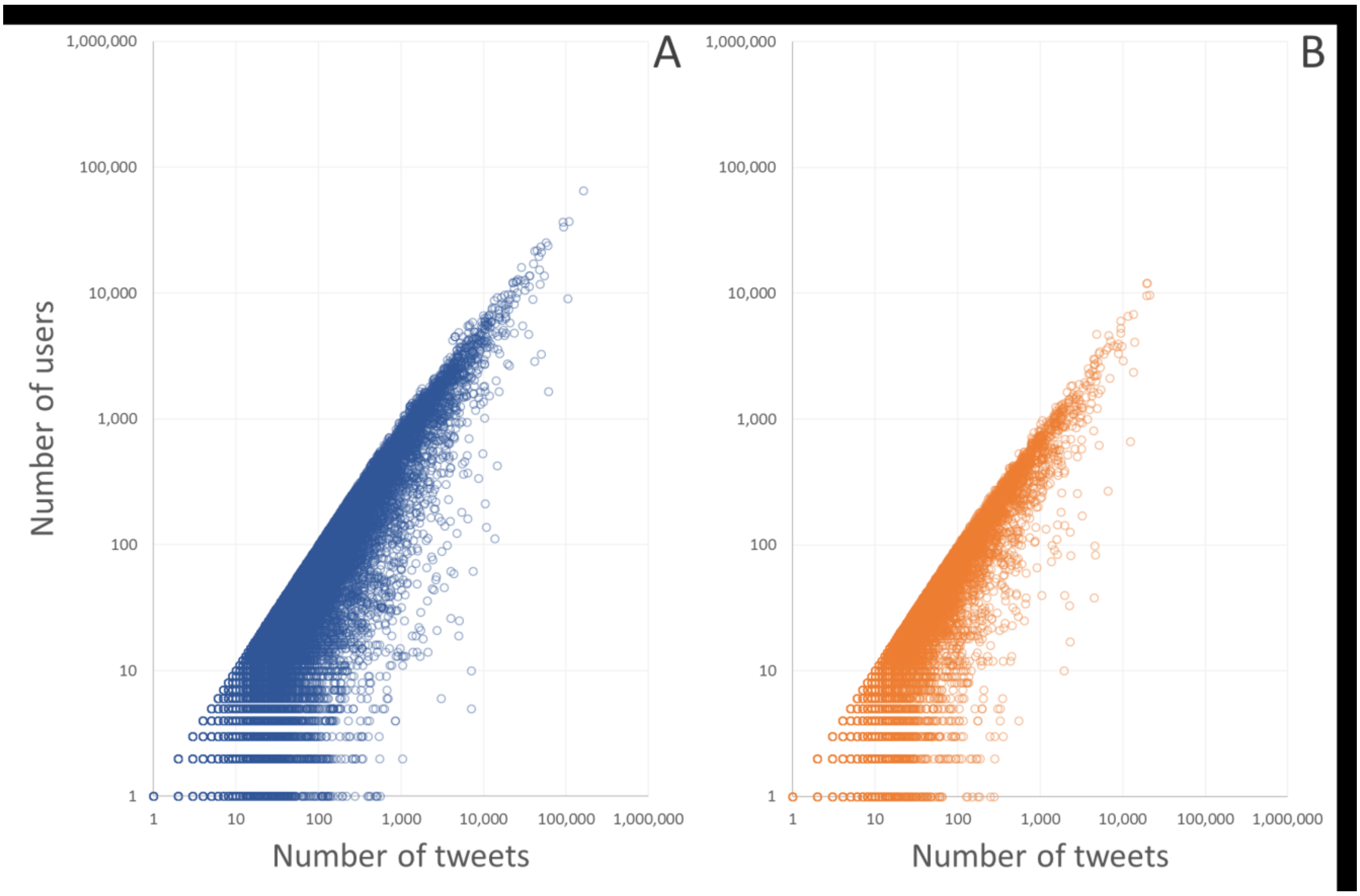

**Figure 3** Number of users and number of tweets per hashtag for all tweets captured by Altmetric (A) and for tweets to 2015 WoS papers (B).

**Table 4** Hashtag statistics for datasets A and B with most frequent hashtags based on number of tweets.

| | **Most frequent hashtags** | **Statistics of hashtag frequency** | Number of tweets | Number of users | Users per hashtag | Number of documents | Number of journals | Tweet span |
|---|---|---|---|---|---|---|---|---|
| Dataset A | #science | Mean | 31 | 17 | 1 | 11 | | 233 |
| | #cancer | *Standard deviation* | *617* | *237* | *4* | *209* | | *459* |
| | #physics | Minimum | 1 | 1 | 1 | 1 | | 0 |
| | #openaccess | Maximum | 162,754 | 65,334 | 1,425 | 54,845 | n/a | 1,825 |
| | #health | 99th percentile | 395 | 229 | 6 | 129 | | 1,746 |
| | #paper | 90th percentile | 20 | 15 | 2 | 8 | | 1,036 |
| | #oa | 75th percentile | 5 | 4 | 1 | 2 | | 182 |
| | | 50th percentile | 2 | 2 | 1 | 1 | | 0 |
| Dataset B | #cancer | Mean | 21 | 13 | 1 | 7 | 4 | 102 |
| | #health | *Standard deviation* | *222* | *111* | *2* | *73* | *19* | *192* |
| | #openaccess | Minimum | 1 | 1 | 1 | 1 | 1 | 0 |
| | #science | Maximum | 21,122 | 12,056 | 274 | 9,765 | 1,684 | 1,723 |
| | #FOAMed | 99th percentile | 311 | 186 | 5 | 92 | 49 | 831 |
| | #Diabetes | 90th percentile | 20 | 15 | 2 | 7 | 6 | 381 |
| | #ornithology | 75th percentile | 6 | 5 | 1 | 2 | 2 | 130 |
| | | 50th percentile | 2 | 2 | 1 | 1 | 1 | 0 |

Table 4 shows hashtag-based stats for both datasets. As to be expected, hashtag frequency and the number of unique terms is greater for dataset A, as it covers all documents in Altmetric and the whole timespan, while dataset B is restricted to WoS documents published in 2015. On average, each hashtag occurred in 21 tweets, was used by 13 users to tag 7 documents and 4 journals indexed in WoS. As shown by the percentiles, hashtag occurrence is extremely skewed. On the individual level, the number of users, documents and journals associated with a hashtag can provide information as to how general and widespread a hashtag is, or how specific and relevant to only a small group of users. The timespan, that is, the number of days between the first and last occurrence of a hashtag, indicates its topicality or timeless relevance. For example, among hashtags that occurred at least 1,000 times, #diet, #water and #nutrition were used during the course of more than four years to describe 2015 documents, while #XmasBMJ lasted only 73 days. The first tweet linking to a 2015 WoS paper with the #diet hashtag appeared 17 November 2011. The discrepancy between tweet date and publication date can be described by the lag between online date and journal issue date (see Haustein et al. (2015b) for an analysis of the publication date problematic).

### 1.3.4 When is scholarly output tweeted?

Tweet activity related to scholarly documents has been shown to occur shortly after publication and disappear within a few days (Eysenbach, 2011; Shuai et al., 2012). Tweeted half-lives and delay between publication and first tweet can thus be measured in hours rather than days. This short-lived attention also points to Twitter being used to diffuse new papers instead of discussing them intensely. It is, however, challenging to accurately calculate delay and decay for all publications in WoS as the publication date of the journal does not sufficiently represent when a publication was actually available. Even with the more accurate article-level information of online dates there are issues to determine the actual date of publication, as demonstrated by tweets mentioning articles before they were supposed to be published (Haustein, Bowman, et al., 2015b). Due to the inaccuracy of available publication dates, tweet delay and tweeted half-lives are not computed.

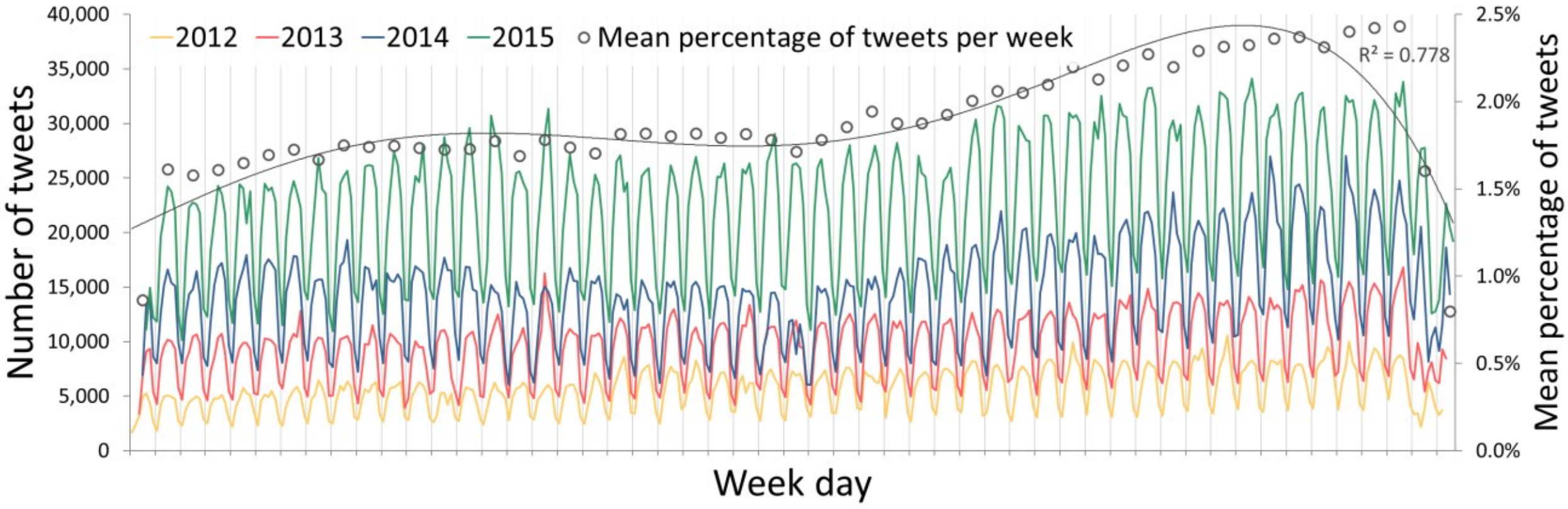

**Figure 4** Number of tweets per week day per year and mean of the weekly percentage per year.

As shown in Figure 4, there are clear differences between weekdays and weekends, reflecting patterns of the work week, which has also been shown to have an effect on journal submissions (Cabanac & Hartley, 2013) and download patterns (Wang et al., 2012). During the week, tweeting activity increases from Monday (14% of tweets) to peak Wednesday (18%) and decrease again towards the weekend. Twitter users tweet, on average, 23% more about scholarly documents on a Wednesday and 41% less on a Sunday. Figure 4 also shows the different magnitude of Twitter activity among years, as well as an overall increase throughout each year. While a general increase from January to December can be observed for each of the four years of tweets, the general trend also reflects the academic year: activity is higher in spring and fall and drops slightly in summer and particularly during the winter break during the last two and first weeks of each year. Considering that Twitter activity often climaxes the day of or day after publication, the season and weekday of publication might influence a document's visibility on Twitter (Haustein, Bowman, et al., 2015b). Zhang and Paxson identified Twitter bots based on tweeting patterns that were too regular to be human (Zhang & Paxson, 2011).

### 1.3.5 Where is scholarly output tweeted?

Twitter provides the possibility to geotag each tweet with precise latitude-longitude information of a user's current location. However, since this function is not activated by default, it is only rarely used. Less than 5% of tweets contain geo coordinates (Graham, Hale, & Gaffney, 2014; Severo, Giraud, & Pecout, 2015), which is why geotags of tweets are not a reliable source to determine where scholarly output is tweeted. Another data source to determine the geographic distribution of Twitter users is to analyze the location information provided in the Twitter bio. However, since the profile location is usually not automatically generated but freely edited by users, it cannot be used without extensive data cleaning. Takhteyec, Gruzd and Wellman (2012) showed that 8% of a sample of 3,360 Twitter profiles contained specific latitude-longitude information, 57% named a location and 20% a country, while 15% used fictional places (e.g., Hogwarts) or too general descriptions to determine the users' whereabouts. Using this profile information, Altmetric is able to determine location information for two-thirds of its tweets. It becomes apparent that in many cases information is not accurate enough to determine the exact location, as remote locations in the UK and Kansas are among the most frequent tweet locations (Haustein & Costas, 2015a).

**Table 5** Top 10 countries by number of users for dataset A and B.

| | | Dataset A | | | | Dataset B | | |
|---|---|---|---|---|---|---|---|---|
| | | **Documents** | **Tweets** | **Users** | | **Documents** | **Tweets** | **Users** |
| Number of unique items | | 3,903,064 | 24,343,105 | 2,622,117 | | 548,841 | 3,960,431 | 601,290 |
| Some country information | | 71% | 58% | 57% | | 70% | 58% | 58% |
| Missing country information | | 69% | 42% | 44% | | 77% | 42% | 42% |
| Top 10 countries by number of users | US | 33.0% | 19.8% | 20.1% | US | 36.2% | 19.5% | 19.4% |
| | GB | 21.6% | 11.0% | 8.3% | GB | 27.4% | 12.3% | 10.6% |
| | CA | 7.9% | 2.9% | 3.1% | CA | 9.3% | 3.1% | 3.4% |
| | JP | 4.3% | 1.8% | 2.3% | ES | 9.6% | 3.3% | 2.7% |
| | AU | 6.5% | 2.9% | 2.3% | AU | 7.6% | 2.5% | 2.4% |
| | ES | 7.5% | 2.8% | 2.2% | JP | 4.4% | 1.5% | 2.0% |
| | FR | 5.3% | 1.5% | 1.4% | NL | 4.0% | 1.0% | 1.3% |
| | IN | 2.6% | 0.7% | 1.1% | FR | 7.9% | 1.6% | 1.2% |
| | NL | 3.5% | 1.0% | 1.1% | DE | 5.6% | 1.1% | 0.9% |
| | DE | 4.1% | 1.0% | 0.9% | IN | 3.4% | 0.7% | 0.9% |

Due to these limitations, the analysis of where users tweet scholarly documents is restricted to the country level (Table 5). Altmetric provides location information for 57% of users, 58% of tweets and 71% of documents (dataset A) and 58% of both users and tweets and 70% of documents for the WoS subset (dataset B). Users from the US are overrepresented as 20% of tweets are sent by Twitter users with an identified location in the US, which is followed by the UK (8%), Canada (3%), Japan, Australia and Spain (2% each). A similar distribution can be observed for the 2015 WoS articles (dataset B), as the top 10 countries by number of users stay the same, although the UK, the Netherlands, Spain, Canada and to a lesser extent Australia and Germany gain in percentage of users, while India, France, Japan and the US lose in comparison to dataset A. While altmetrics have been marketed as democratizers of science evaluation in the terms of having the potential to correct for biases created by WoS and other US and English-centric journal databases, these results show that, when it comes to Twitter, known biases persist or are even intensified on social media.

### 1.3.6 Who tweets scholarly output?

One of the main motivations to consider tweets as an altmetric indicator is that Twitter is used by the general public and thus, at least theoretically, offers insight into how non-academics engage with scholarly output. In order to separate tweets by the public from those sent by members of the scholarly community, Twitter users have to be identified and classified as such.

#### *1.3.6.1 Identifying users who tweet scholarly content*

One of the main challenges of determining the type of impact reflected by tweets to scientific papers is to identify who is tweeting. While Mendeley provides certain standardized user demographics such as academic status, discipline or country for users associated with a paper, the classification of tweets by user type is restricted to Twitter bios. These self-descriptions are 160-character texts, which provide users with the space to present themselves to other users of the microblogging platform.

Applying a codebook to determine who tweets scientific papers based on Twitter username, bio and photo, a sample of 2,000 accounts tweeting links to articles published in *Nature*, *PLOS ONE*, *PNAS* and *Science*, Tsou et al. (2015) found that almost one quarter of accounts were maintained by an organization. Among these were mainly non-profits (42%), corporations (29%) and universities (13%), while many were also classified as news, media or outreach institutions (19%). Among the 1,520 accounts identified as individuals, two thirds were male. One third of the users were identified as having a PhD and 12% as students. This amounts to almost half of all identified individuals having completed or pursuing a doctorate degree, which stands in glaring contrast to about 1% of the US population with a PhD (Tsou et al., 2015), strongly suggesting that it is the scholarly community rather than the general public who tweets links to scientific papers.

Applying a similar codebook to a random sample of 800 accounts tweeting 2012 WoS papers, 68% of accounts were maintained by an individual, 21% by an organization, while 12% could not be identified (Haustein et al., 2016), corroborating the findings by Tsou et al. (2015). Among individuals, 47% used professional terms (e.g., doctor, MD, photographer) to describe themselves, 22% identified as researchers (e.g., scientist, professor, postdoc), 13% as science communicators (e.g., writer, author, journalist, blogger) and 7% as students (e.g., grad student, PhD candidate). Reflecting the blurred boundaries between personal and professional communication, many individuals used words from more than one of these categories to describe themselves. For example, 8% of accounts were classified as researchers and professionals and 5% as professionals and science communicators (Haustein et al., 2016). Science communicators were also the largest group of Twitter users mentioned by astrophysicists (Holmberg et al., 2014). Although labor-intensive and based on little more than 160-character self-descriptions, the above studies show that it is feasible to extract members of academia from users tweeting scholarly documents. Keyword-based searches can be applied to identify scholars in larger samples, but are limited by either low recall or low precision depending on the particular query (Barthel et al., 2015).

It is considerably more challenging to identify members of the general public. Although many Twitter bios contain terms depicting personal lives (e.g., father, wife, yoga lover), the presence of these terms does not necessarily mean that accounts are maintained by non-academics, because scholars often describe themselves in both personal and professional manner on Twitter (Bowman, 2015b; Haustein & Costas, 2015b; Work et al., 2015). Similarly, it is challenging to distinguish members of the public based on an indeterminate list of terms of non-academic professions (e.g., consultant, photographer), especially when also considering accounts in languages other than English. Even the comparably straight-forward identification of a researcher who strictly identifies as such on Twitter, becomes problematic when they shared a paper out of private interests. For example, a tweet by a physicist might actually reflect engagement by the public rather than scholarly communication, if they tweeted about a cancer study as a member of a patient group rather than in their academic role.

An alternative to classifying users based on publicly available information, is to approach them directly and ask them who they are. Alperin (2015b) pioneered such a survey method on Twitter, which with the help of an automated Twitter account, asked users who had tweeted a *Scielo Brazil* paper, whether they were affiliated with a university. Such a direct approach might also be helpful to determine the motivation of a user to tweet a specific paper, helping to quantify and distinguish different types of tweets, such as endorsement or critical discusion, diffusion or self-promotion. Author self-citations or self-tweets accounted for 7% of a sample of 270 tweets (Thelwall et al., 2013).

**Table 6** Number of followers, tweets, tweet span, tweets per day and tweeting activity per week for the most active users in dataset A (≥20,000 tweets per user).

| Twitter handle | Number of followers | Tweets | Tweet span | Tweets per day | Tweeting activity per week |
|---|---|---|---|---|---|
| blackphysicists | 12,914 | 92,583 | 1,826 | 50.7 | |
| MathPaper | 1,889 | 73,239 | 1,086 | 67.4 | |
| anestesiaf | 1,428 | 63,953 | 1,713 | 37.3 | |
| oceanologia | 389 | 62,585 | 1,353 | 46.3 | |
| UIST_Papers20XX | 35 | 50,211 | 1,360 | 36.9 | |
| UIST_Papers19XX | 6 | 49,838 | 1,359 | 36.7 | |
| hiv_insight | 14,328 | 48,714 | 1,822 | 26.7 | |
| russfeed | 2,127 | 44,463 | 577 | 77.1 | |
| uranus_2 | 2,519 | 42,053 | 1,790 | 23.5 | |
| Immunol_papers | 477 | 40,721 | 639 | 63.7 | |
| psych2evidence | 513 | 40,658 | 410 | 99.2 | |
| InorganicNews | 1,044 | 36,869 | 1,657 | 22.3 | |
| arXiv_trend | 166 | 33,272 | 457 | 72.8 | |
| hlth_literacy | 5,582 | 32,337 | 1,823 | 17.7 | |
| AstroPHYPapers | 3,427 | 31,154 | 1,086 | 28.7 | |
| cirugiaf | 406 | 30,692 | 1,710 | 17.9 | |
| ThihaSwe_dr | 601 | 29,357 | 1,164 | 25.2 | |
| semantic_bot | 0 | 28,908 | 28 | 1032.4 | |
| CondensedPapers | 530 | 27,603 | 912 | 30.3 | |
| libroazuln | 128 | 24,994 | 1,613 | 15.5 | |
| rnomics | 1,520 | 24,905 | 1,821 | 13.7 | |
| PhysicsPaper | 476 | 22,894 | 1,086 | 21.1 | |
| epigen_papers | 788 | 21,916 | 739 | 29.7 | |

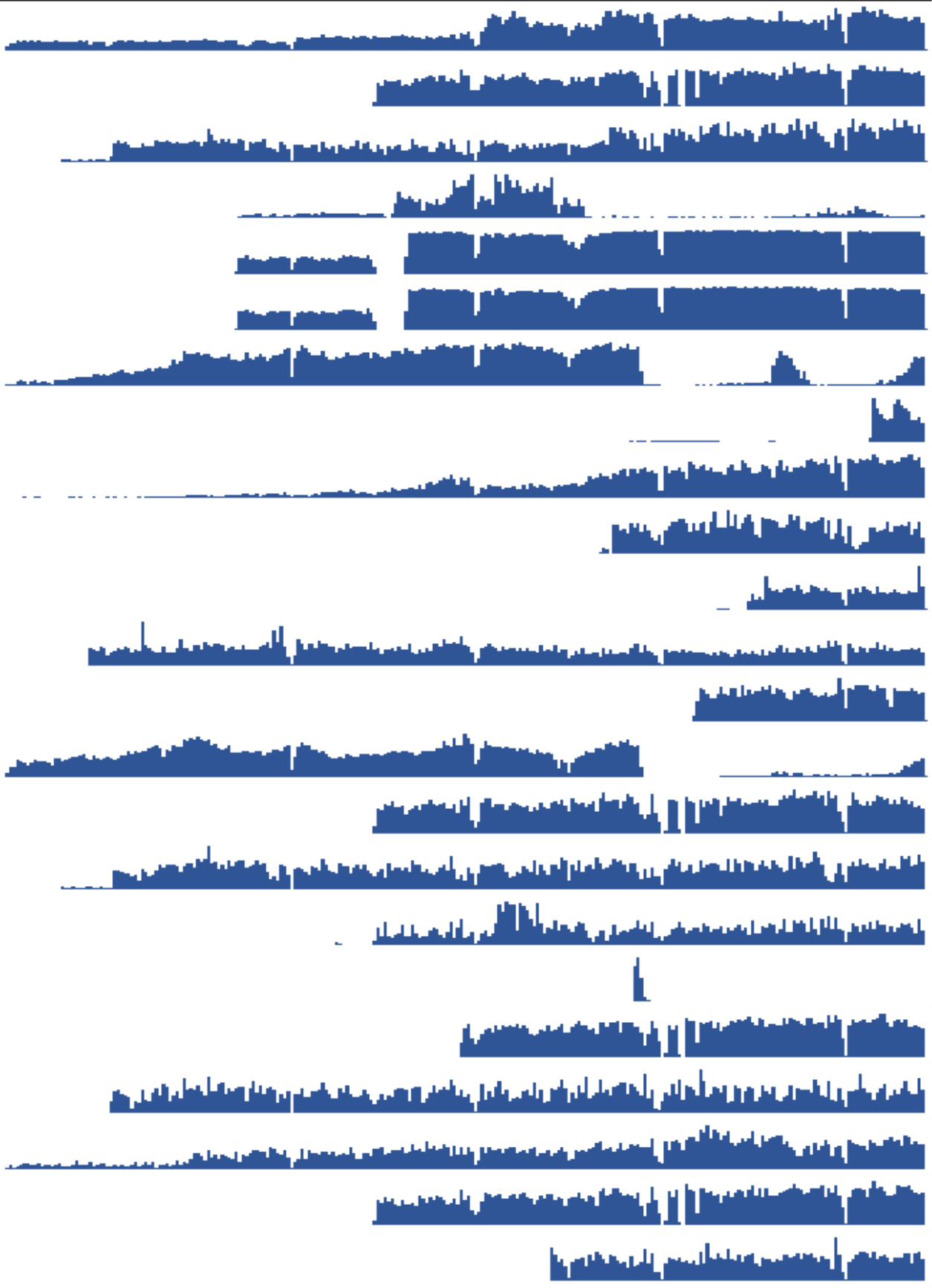

The 24.3 million tweets captured by Altmetric were sent by 2.6 million users. Looking at the most active users who tweeted more than 1,000 times during the whole period covered by Altmetric (Table 6), the presence of accounts automatically diffusing scholarly articles on Twitter becomes apparent (Haustein, Bowman, Holmberg, et al., 2016). In fact, 15 of the 19 most productive accounts in Table 6 with more than 25,000 tweets self-identified bots (see below).

### *1.3.6.2 Classifying users by Twitter activity*

Instead of classifying users according to their self-descriptions, accounts can also be grouped based on their activity. Dividing Twitter accounts into three groups of top 1%, 9% and 90% of users (according to number of tweets) helps to distinguish lead and highly active users from less active ones (Bruns & Stieglitz, 2014). This classification provides insights into tweeting behavior of different types of users. Separating the 601,290 users in dataset B by number of tweets linking to a 2015 WoS article, 6,016 lead users, 54,535 highly active and 540,739 least active users can be identified. Lead users contributed between 84 and 19,973 tweets, a median of 149 tweets per users, had on average 935 followers (median; mean=3,862) and tweeted the 2015 papers during an average tweet span of 598 days. Highly active users contributed between 9 and 83 tweets (median=16), had less followers (median=442.5; mean=2,136) and shorter tweet spans (mean=388 days), while least active users tweeted up to eight times (median=1), had 212 followers and were active for a period of 58 days.

Lead (top 1% of users), highly active (9%) and least active (90%) users contributed 43%, 31% and 25% of tweets to the entire set of 2015 WoS papers, respectively (Figure 5). Interestingly, these percentages differ among NSF disciplines with least active users overrepresented among those tweeting literature from Professional Fields, Social Sciences, Psychology and Earth and Space. On the contrary, lead users were overrepresented in Chemistry, Physics, Mathematics and Engineering & Technology, which were the fields exhibiting the lowest Twitter coverage, density and number of unique users (Table 2). Assuming that the general public is least active when it comes to tweeting about scholarly papers, they are more likely to engage with articles published in journals from the Professional Fields and Social Sciences and less likely to tweet Chemical papers. The high presence of lead users in Chemistry and Physics might, at least partly, be caused by accounts promoting these papers automatically, such as @blackphysicists and @MathPaper, which were the two most active accounts in dataset A, tweeting 51 and 67 scholarly documents per day (Table 6).

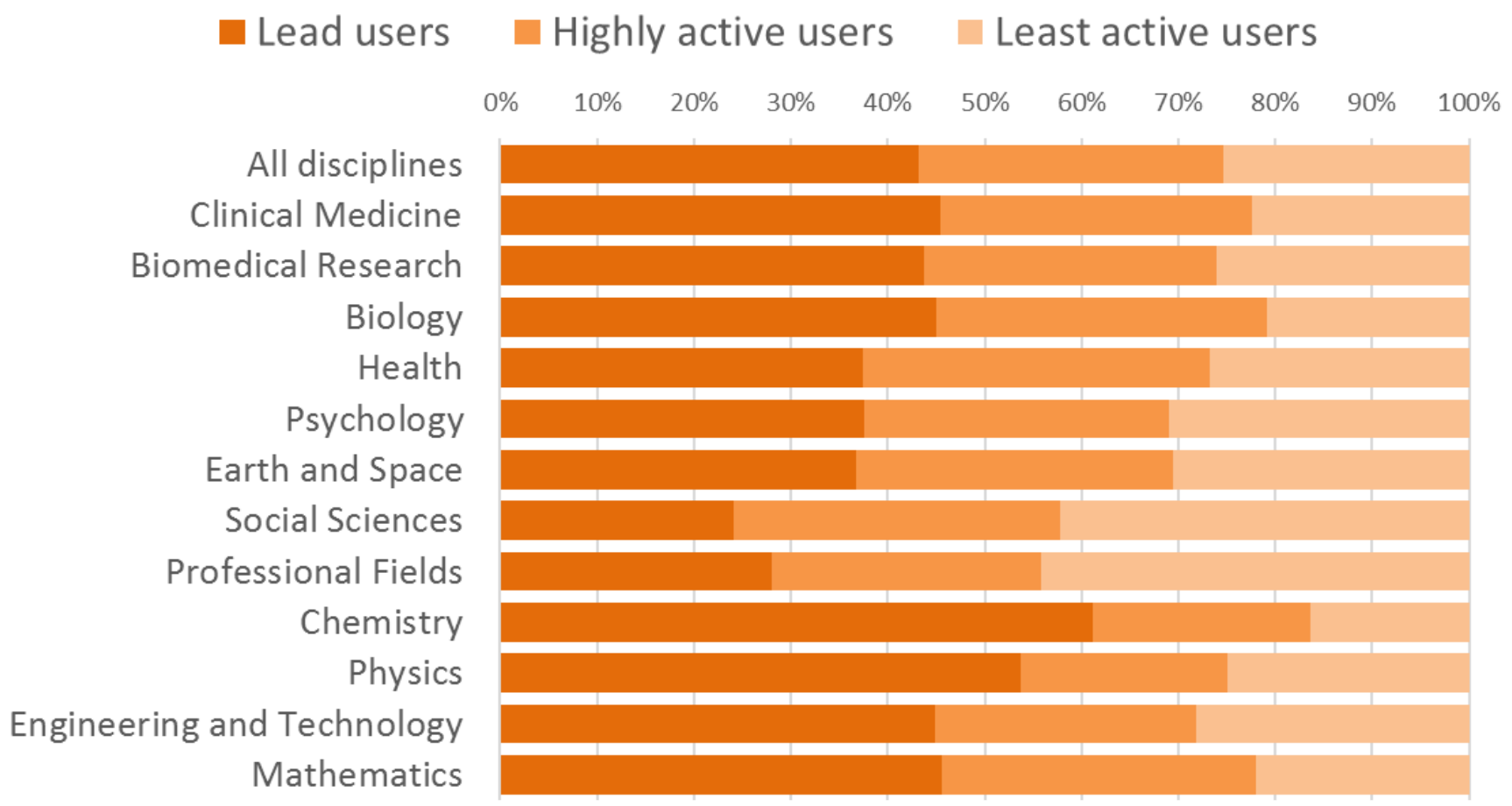

**Figure 5** Percentage of tweets from lead users (1%), highly active users (9%) and least active users (90%) per discipline (dataset B).

### *1.3.6.3 Twitter bots*

Automated Twitter accounts have become prevalent on Twitter (Mowbray, 2014). About one fifth of tweets sent during the 2016 presidential election were estimated to be sent by bots (Bessi & Ferrara, 2016) and almost one quarter of tweets in 2009 came from accounts tweeting more than 150 times per day. Ferrara et al. (2016, p. 96) define social bots as "a computer algorithm that automatically produces and interacts with humans on social media, trying to emulate and possibly alter their behavior". Automated Twitter accounts can be further distinguished between useful bots and antisocial or spambots (Ferrara et al., 2016; Mowbray, 2014).

Bots are also infiltrating academic Twitter. Among a random sample of 800 Twitter accounts captured by Altmetric, 8% seemed completely and 5% partially automated (Haustein et al., 2016), while automated accounts who self-identified as such were responsible for 9% of tweets to *arXiv* submissions (Haustein, Bowman, Holmberg, et al., 2016). Shuai et al. (2012) even removed half of the tweets to a sample of tweets to documents on *arXiv*, as they were created by bots. Regardless of whether Twitter accounts that automatically tweet scientific papers are considered useful or spam, it is safe to say that their automated tweets do not reflect *impact*. In the context of altmetrics and tweets to scientific papers, bot activity thus needs to be at least identified, if not entirely removed in an impact assessment. Although spammers and excessive self-promotion was identified as a challenge by Altmetric, they still considered gaming a rare and easy to identify threat in 2013 (Liu & Adie, 2013).

Twitter's terms of service specifically address automation. While prohibiting spam, Twitter encourages automated tweets if they "broadcast helpful information" (Twitter, 2017). However, what is considered spam continuously evolves as users apply "new tricks and tactics" (Twitter, n.d.) to adapt to or circumvent Twitter rules. Twitter bots can be identified based on specific regularities in their tweeting behavior, such as the frequent and repetitive use of the same hashtags, URLs, tweet format and content, regular temporal activity, as well as the follower/friends ratio, @mentions to non-followers or account suspensions (Mowbray, 2014). Sixteen percent of Twitter accounts were identified as automated based on not-uniform-enough or too-uniform tweeting patterns to be stemming from a human (Zhang & Paxson, 2011). Social network indicators based on the Twitter follower and friend network are considered more robust measures, as they are harder to influence (Yang, Harkreader, & Gu, 2011). The *BotOrNot* algorithm additionally considers linguistic features and tweet sentiment to detect automated Twitter accounts (Davis, Varol, Ferrara, Flammini, & Menczer, 2016). However, as these spamming measures get known and integrated by Twitter to block spam accounts, bots employ more sophisticated algorithms to avoid getting caught, resulting in an arms race between those who create and those who seek to identify Twitter spam.

```
(journal AND NOT (journals OR journalism)) OR jrnl OR publisher
OR publishing OR university press OR Oxford journals OR Cell
Press OR CellPress OR Dove Press OR Taylor & Francis OR CRC
Press OR PortlandPress OR Routledge OR Springer OR Elsevier OR
Wiley OR Wolters Kluwer OR SAGE Publishing
```

**Figure 6** Query used to identify automated Twitter accounts.

```
(robot AND NOT robotics) OR (bot AND NOT (botany OR robotics))
OR (paper OR publication OR lit OR preprint OR article OR peer-
review OR journal) AND feed) OR news feed OR datafeed OR RSS OR
new submissions OR (new AND paper) OR latest publication OR new
publication OR arxiv OR (PubMed AND NOT Chief Editor) OR
bioRxiv OR (papers AND (auto OR stream OR tweet OR updates OR
links)) OR publication alert OR daily updates
```

**Figure 7** 6 Query used to identify publisher accounts.

Analyzing the most active users who have tweeted at least 1,000 times (dataset A), a keyword-based query searching the Twitter bio as well as user name and handle revealed that among 2,043 accounts 248 identified themselves as automated (Figure 6) and 305 as journal or publisher accounts (Figure 7). These make up 30% and 11% of the tweets sent by the 2,043 most active users (Table 7), which correspond to 7% and 3% of the entire 24.3 million tweets to scholarly documents in dataset A. The median number of followers is significantly lower than other accounts (Table 7) and as few as 6% of the 1.8 million tweets sent by the 248 accounts contain @mentions. If scholarly bots do mention other users, they often seem to reference journals such as @hiv_insight, which frequently mentions @PLoSMedicine, @STI_BMJ and @JAMA_current. Other than social bots in the general Twittersphere, scholarly bots seem to not try to emulate human behavior or game the system. They rather resemble RSS feeds tweeting the paper title and a link, often specifying what type of information they diffuse. Some even provide instructions to create similar feeds. For example, the Twitter bio of @asthma_papers reads "RSS feed for #asthma papers in #Pubmed. Create a feed of your own using instructions here: https://github.com/roblanf/phypapers".

**Table 7** Twitter metrics for self-identified bots, journal and publisher accounts and other accounts based on users with at least 1,000 tweets (dataset A).

| **Most active Twitter accounts** (≥1,000 tweets, n=2,043) | | **Followers** | **Tweets** | **Tweets per day** | **Tweet span** |
|---|---|---|---|---|---|
| Self-identified bots n=248 30% of tweets | median | 212 | 3,479 | 5.1 | 845 |
| | mean | 1,014 | 7,339 | 14.4 | 923 |
| | *std dev* | *2,781* | *10,390* | *67.0* | *477* |
| | min | 0 | 1,001 | 0.6 | 28 |
| | max | 25,003 | 73,239 | 1,032.4 | 1,823 |
| Journal and publisher accounts n=305 11% of tweets | median | 3,199 | 1,670 | 1.2 | 1,647 |
| | mean | 21,475 | 2,249 | 1.7 | 1,484 |
| | *std dev* | *116,124* | *1,874* | *1.6* | *369* |
| | min | 3 | 1,001 | 0.6 | 122 |
| | max | 1,448,649 | 19,256 | 14.7 | 1,822 |
| Other accounts n=1,490 59% of tweets | median | 1,535 | 1,599 | 1.4 | 1,388 |
| | mean | 5,236 | 2,408 | 2.9 | 1,278 |
| | *std dev* | *14,091* | *3,670* | *9.3* | *496* |
| | min | 0 | 1,000 | 0.6 | 8 |
| | max | 228,224 | 92,583 | 297.3 | 1,826 |

Considering that scholars use Twitter to diffuse information and stay aware of relevant literature, automated accounts might be considered useful. However, bots have shown to be harmful to society, when they are used to influence public opinion and behavior such as political opinions or elections (Bessi & Ferrara, 2016; Metaxas & Mustafaraj, 2012; Ratkiewicz, Conover, Meiss, Gonçalves, Flammini, et al., 2011; Ratkiewicz, Conover, Meiss, Gonçalves, Patil, et al., 2011) or manipulation of the stock market (Ferrara et al., 2016). If Twitter impact became part of the scholarly reward system, Twitter bots might be able to similarly influence opinions or shape outcomes of certain research metrics.

## 1.4 Conclusion and outlook

This chapter provided an overview of the use of Twitter in scholarly communication. By demonstrating *who* uses Twitter in academia and for what reasons, *what* types of scholarly outputs are diffused *how*, *where* and *when*, it aimed to add context and help to interpret any scholarly metrics derived from this and similar types of social media activity.

Research evaluators and managers were particularly excited at the prospect of an easily accessible data source that would be able to capture traces of the societal impact of research. However, perhaps unsurprisingly, the majority of tweets stem from stakeholders in academia rather than from members of the general public, which indicates that the majority of tweets to scientific papers are more likely to reflect scholarly communication rather than societal impact. At the same time, Twitter uptake in academia lacks behind Twitter user by the general public. Twitter activity is influenced by geographical and disciplinary biases and publication date. Known biases towards US and UK sources persist, rather than democratizing scholarly communication and the reward system of science.

The majority of tweets linking to scientific articles appear shortly after their publication; tweeting half-lives can be measured in hours rather than days. Moreover, one can observe weekday as well as seasonal patterns with Twitter activity peaking Wednesdays and in the fall and plummeting during the weekend and holiday season. Journal and publisher accounts as well as Twitter bots contribute significantly to tweeting activity linked to academic papers, which suggests that a significant extent of tweeting activity serves promotional purposes or is automated, which does neither reflect societal nor scholarly impact. A large share of tweets contains hashtags and mention either the title or a short summary of the paper they referred to. Half of all articles linking to 2015 WoS papers were retweets and the majority contained no sentiments. These tweeting characteristics emphasize particular low engagement of users linking to journal articles. The main motivation for researchers to use Twitter is information diffusion, networking and to stay up-to-date with the literature. However, the sheer brevity of tweets makes intense discussions the exception rather than the rule on Twitter.

As citation behavior and motivations to cite or not cite certain sources are biased and influenced by many factors other than a paper's significance, not every citation represents impact. However, each scholarly author is bound by scholarly norms to participate in the citation process. In some rare cases, where scholars tweet corrections to publications or journals provide tweetable abstracts and organize journal clubs, Twitter has started to be integrated in or even replace certain functions of formal journal publishing. However, in most fields tweeting does not yet play an important role in scholarly communication. In most disciplines, Twitter uptake is low and the platform is only used in a passive or infrequent manner, or tweets reflect only a part of informal scholarly communication, such as conference chatter. Moreover, Twitter uptake varies between disciplines, countries, journals, as well as individuals and can be easily influenced and manipulated. The presence of automated Twitter accounts which promote certain contents becomes particularly problematic when tweet counts become the basis for measures of impact.

This is not to say that Twitter should be completely disregarded as a data source for scholarly metrics. Rather, the microblogging platform should be approached critically in terms of what kind of use and user populations it captures. By reviewing the role of Twitter in scholarly communication and analyzing tweets linking to scholarly documents in depths and beyond crude counts, this chapter attepmted to provide the basis for more sophisticated and well-balanced approaches to scholarly Twitter metrics.